\documentclass[11pt]{article}
\usepackage{amssymb,amsmath,amsfonts}
\usepackage{graphicx}
\usepackage{graphics}
\usepackage{eepic,epsfig}

\makeatletter \@addtoreset{equation}{section}

\makeatother

\begin{document}

\title{Wightman function and the Casimir effect for \\
a Robin sphere in a constant curvature space}
\author{ S. Bellucci$^{1}$\thanks{%
E-mail: bellucci@lnf.infn.it }, A. A. Saharian$^{2}$\thanks{%
E-mail: saharian@ysu.am }, N. A. Saharyan$^{2}$ \vspace{0.3cm} \\
\textit{$^1$ INFN, Laboratori Nazionali di Frascati,}\\
\textit{Via Enrico Fermi 40, 00044 Frascati, Italy} \vspace{0.3cm}\\
\textit{$^2$ Department of Physics, Yerevan State University,}\\
\textit{1 Alex Manoogian Street, 0025 Yerevan, Armenia }}
\maketitle

\begin{abstract}
We evaluate the Wightman function, the mean field squared and the vacuum
expectation value (VEV) of the energy-momentum tensor for a scalar field
with Robin boundary condition on a spherical shell in the background of a
constant negative curvature space. For the coefficient in the boundary
condition there is a critical value above which the scalar vacuum becomes
unstable. In both interior and exterior regions, the VEVs are decomposed
into the boundary-free and sphere-induced contributions. For the latter,
rapidly convergent integral representations are provided. In the region
inside the sphere, the eigenvalues are expressed in terms of the zeros of
the combination of the associated Legendre function and its derivative and
the decomposition is achieved by making use of the Abel-Plana type summation
formula for the series over these zeros. The sphere-induced contribution to
the VEV of the field squared is negative for Dirichlet boundary condition
and positive for Neumann one. At distances from the sphere larger than the
curvature scale of the background space the suppression of the vacuum
fluctuations in the gravitational field corresponding to the negative
curvature space is stronger compared with the case of the Minkowskian bulk.
In particular, the decay of the VEVs with the distance is exponential for
both massive and massless fields. The corresponding results are generalized
for spaces with spherical bubbles and for cosmological models with negative
curvature spaces.
\end{abstract}

\bigskip


\bigskip

\section{Introduction}

\label{sec:Introd}

In the absence of a reliable theory for quantum gravity, the influence of
the gravitational field on quantum matter is investigated within the
framework of semiclassical theory. In the latter, the gravitational field is
considered as the classical background and the back reaction of quantum
effects is described by quasiclassical Einstein equations with the
expectation value of the energy-momentum tensor for quantum fields in the
right-hand side (for reviews see \cite{Birr82}-\cite{Mukh09}). Among the
most interesting effects in this field are the particle production and the
vacuum polarization by strong gravitational fields. A well known example is
the thermal radiation emitted by black holes.

The investigation of quantum effects in curved backgrounds is motivated by
several reasons. During the cosmological expansion of the Universe, the back
reaction of the particles created by a time dependent gravitational field at
early stages leads to a rapid isotropization of the expansion. According to
the inflationary scenario, the quantum vacuum fluctuations of a scalar
field, amplified by the gravitational field during the quasi-de Sitter like
expansion, serve as seeds for the large scale structure formation in the
Universe \cite{Lind90}. The information on the properties of these
fluctuations are encoded in the thermal anisotropies of the cosmic microwave
background which are measured with high accuracy by a number of recent
cosmological projects. An important feature of quantum field theory in
curved backgrounds is a possible breakdown of the energy conditions in the
formulations of Hawking-Penrose singularity theorems, by the expectation
value of the energy-momentum tensor of quantum fields. This opens a
possibility to solve the singularity problem within the framework of
classical general relativity. An additional motivation for the study of
quantum field theoretical effects in curved backgrounds appeared recently in
condensed matter physics related to various analog models. In particular,
the long wavelength properties of the electronic subsystem in a graphene
sheet are well described by the Dirac-like model with the speed of light
replaced by the Fermi velocity. In curved graphene structures (for example,
in fullerenes) the corresponding field theory is formulated on curved
geometry and the curvature effects should be taken into account in the
calculations of physical properties of these structures \cite{Cort12,Iori13}.

A large number of problems, in addition to the curvature of the background
spacetime, involve boundaries on which the operator of a quantum field obeys
prescribed boundary conditions. The physical nature of the boundaries can be
different. The examples are macroscopic bodies in QED, edges of graphene
sheets in nanoribbons, interfaces separating different phases of a physical
system, horizons in gravitational physics, the branes in higher-dimensional
field-theoretical models. Among the most interesting macroscopic
manifestations of the influence of boundaries on the properties of a quantum
field is the Casimir effect (see Refs. \cite{Most97,Milt02}). It arises as a
consequence of the modification of the quantum fluctuations spectrum by the
boundary conditions imposed on the field. As a result of this, the
expectation values of physical observables are changed and forces arise
acting on the constraining boundaries. The effect for scalar, fermionic and
vector fields is investigated in various bulk and boundary geometries and
for various types of boundary conditions.

An interesting topic in the investigations of the Casimir effect is the
dependence of the physical characteristics of the vacuum, like the energy
density and stresses, on the geometry of the background spacetime. The
evaluation of these characteristics requires the knowledge of a complete set
of modes for a quantum field and exact results can be provided for highly
symmetric bulk and boundary geometries only. In particular, motivated by the
problems of the radion stabilization and the cosmological constant
generation in braneworld models, the Casimir effect for planar boundaries in
anti-de Sitter space has been widely discussed \cite{Gold00}.
Higher-dimensional generalizations of the anti-de Sitter background having
compact internal spaces have been considered as well \cite{Flac03}. Another
class of exactly solvable Casimir problems corresponds to de Sitter
spacetime. Various geometries of boundaries in this background have been
considered \cite{Seta01}. In inflationary coordinates the corresponding
metric is time-dependent and, in general, in addition to the diagonal
component, the vacuum energy-momentum tensor has a nonzero off-diagonal
component that describes an energy flux along the direction normal to the
boundary. The Casimir effect for the electromagnetic field in the geometry
of parallel conducting plates in Friedmann-Robertson-Walker cosmologies with
flat spatial sections and with power-law scale factors is discussed in \cite%
{Bell13}.

In the present paper we consider the change in the properties of the scalar
vacuum induced by a spherical shell with Robin boundary condition on the
background of a negatively curved constant curvature space. Historically,
the investigation of the Casimir effect with a spherical boundary was
motivated by the semiclassical model of an electron \cite{Casi53}, where the
vacuum quantum fluctuations of the electromagnetic field are responsible for
Poincar\'{e} stress stabilizing the charged particle. Currently the
configurations involving spherical boundaries are among the most popular
ones in both the theoretical and experimental investigations of the Casimir
effect (see \cite{Most97,Milt02} and references therein). Here, we are
interested in combined effects of the gravitational field and spherical
boundary on the local characteristics of the scalar vacuum. We shall
consider a free field theory and in this case all the properties of the
vacuum state are obtained from two-point functions. Our choice of the
positive-frequency Wightman function is motivated by the fact that this
function will also determine the response of a Unruh--De Witt detector (see,
for instance, \cite{Birr82}). As local characteristics of the vacuum state
we shall investigate the vacuum expectation values (VEVs) of the field
squared and the energy-momentum tensor. The latter serves as a source in the
right-hand side of semiclassical Einstein equations and plays an important
role in considerations of the back-reaction from quantum effects on the
background geometry.

The paper is organized as follows. In Section \ref{sec:WF}, we describe the
bulk and boundary geometries. The positive-frequency Wightman function is
evaluated for the boundary-free space, inside and outside a spherical
boundary. By making use of this function, in Section \ref{sec:phi2}, the
sphere-induced contribution in the VEV of the field squared is evaluated and
its properties are investigated in asymptotic regions of the parameters. The
VEV of the energy-momentum tensor is studied in Section \ref{sec:EMT} for
the both interior and exterior regions. In Section \ref{sec:Bubble}, we
consider the background spacetime with the geometry described by distinct
metric tensors inside and outside a spherical boundary. Two cases are
investigated. In the first one the interior geometry is described by a
general spherically symmetric static metric and the exterior metric
corresponds to a constant negative curvature space. An example is considered
with an interior Minkowskian geometry. In the second case, a constant
curvature space is realized in the interior region whereas the exterior
geometry is the Minkowski one. The main results are summarized in Section %
\ref{sec:Conclus}. For the interior geometry with a negative constant
curvature space, the eigenmodes of a quantum scalar field with Robin
boundary condition are expressed in terms of the zeros of the associated
Legendre function with respect to its order. In appendix \ref{sec:Zeros}, we
discuss the properties of these zeros. The expression of the Wightman
function inside the spherical shell contains the summation over these zeros.
In appendix \ref{sec:SF}, by making use of the generalized Abel-Plana
formula, a summation formula is derived for the series of this type.

\section{Wightman function}

\label{sec:WF}

\subsection{Geometry of the problem}

We consider a quantum scalar field $\varphi (x)$ with the curvature coupling
parameter $\xi $. The most important special cases of this parameter $\xi =0$
and $\xi =\xi _{D}=(D-1)/(4D)$ correspond to minimally and conformally
coupled scalars, respectively. The field equation has the form%
\begin{equation}
(\nabla _{l}\nabla ^{l}+m^{2}+\xi \mathcal{R})\varphi (x)=0,  \label{fieldEq}
\end{equation}%
where $\mathcal{R}$ is the scalar curvature for the background spacetime.
The background geometry is described by the line element
\begin{equation}
ds^{2}=dt^{2}-a^{2}(dr^{2}+\sinh ^{2}rd\Omega _{D-1}^{2}),  \label{ds2}
\end{equation}%
with a constant $a$ and with $d\Omega _{D-1}^{2}$ being the line element on
the $(D-1)$-dimensional sphere, $S^{D-1}$. The corresponding angular
coordinates we denote by $(\vartheta ,\phi )$ with $\vartheta =(\theta
_{1},\ldots ,\theta _{n})$, $n=D-2$, and $0\leqslant \theta _{k}\leqslant
\pi $, $k=1,\ldots ,n$, $0\leqslant \phi \leqslant 2\pi $. The spatial part
of the line element (\ref{ds2}) describes a constant negative curvature
space. The spaces with negative curvature play a significant role in
cosmology and in holographic theories. For the Ricci scalar corresponding to
(\ref{ds2}) one has $\mathcal{R}=-D(D-1)/a^{2}$. Note that in Eq. (\ref%
{fieldEq}) the curvature coupling term appears in the form of the effective
mass squared $m^{2}-D(D-1)\xi /a^{2}$. Depending on the value of the
curvature coupling parameter, the latter can be either negative or positive.

Quantum effects on the background of constant curvature spaces have been
widely discussed in the literature (see, for instance, \cite%
{Birr82,Grib94,Most97,Milt02,Camp90}). These effects play an important role
in the physics of the early Universe, in inflationary models, and in a
number of condensed matter systems described by effective curved geometries.
Here we are interested in combined effects of the background gravitational
field and boundaries on the properties of the quantum vacuum for a scalar
field. As a boundary geometry we consider a spherical shell on which the
field operator obeys Robin boundary condition (BC)%
\begin{equation}
\left( A+Bn^{l}\nabla _{l}\right) \varphi (x)=0,\;r=r_{0},  \label{Robin}
\end{equation}%
where $r_{0}$ is the sphere radius, $A$ and $B$ are constants, and $n^{l}$
is the unit inward normal to the sphere, $n^{l}=-\delta _{(j)}\delta
_{1}^{l}/a$, $j=i,e$, with $\delta _{(i)}=1$ for the interior region and $%
\delta _{(e)}=-1$ for the exterior region. With this, for $B\neq 0$, the BC
can also be written in the form $\left( \beta -\delta _{(j)}\partial
_{r}\right) \varphi (x)=0$, $r=r_{0}$, with the notation
\begin{equation}
\beta =aA/B.  \label{beta}
\end{equation}%
Of course, all the physical results will depend on this ratio. We wrote the
condition in the form (\ref{Robin}) to keep the transition to special cases
of Dirichlet ($B=0$) and Neumann ($A=0$) BCs transparent.

For a free field theory all the properties of the quantum vacuum are
contained in two-point functions. Here we shall consider the
positive-frequency Wightman function defined as the VEV $W(x,x^{\prime
})=\langle 0|\varphi (x)\varphi (x^{\prime })|0\rangle $, where $|0\rangle $
stands for the vacuum state. The VEVs of physical characteristics bilinear
in the field operator, such as the field squared and the energy-momentum
tensor, are obtained from this function in the coincidence limit. In
addition to this, the positive-frequency Wightman function determines the
response function for the Unruh--De Witt particle detector in a given state
of motion \cite{Birr82}.

Let $\{\varphi _{\alpha }(x),\varphi _{\alpha }^{\ast }(x)\}$ be a complete
set of normalized positive- and negative-energy mode functions obeying the
field equation (\ref{fieldEq}) and the BC (\ref{Robin}). Here, the
collective index $\alpha $ is the set of quantum numbers specifying the
solutions. For the problem under consideration, the positive-energy mode
functions can be presented in the factorized form%
\begin{equation}
\varphi _{\alpha }(x)=R_{l}(r)Y(m_{k};\vartheta ,\phi )e^{-iEt},
\label{eigfunc1}
\end{equation}%
where $Y(m_{k};\vartheta ,\phi )$ is the spherical harmonic of degree $l$.
For the angular quantum numbers one has $l=0,1,2\ldots $, $%
m_{k}=(m_{0}=l,m_{1},\ldots ,m_{n})$, where $m_{1}$, $m_{2}$,..., $m_{n}$
are integers obeying the relations
\begin{equation}
0\leqslant m_{n-1}\leqslant m_{n-2}\leqslant \cdots \leqslant m_{1}\leqslant
l,  \label{mrange}
\end{equation}%
and $-m_{n-1}\leqslant m_{n}\leqslant m_{n-1}$. By taking into account the
equation for the spherical harmonics,
\begin{equation}
\Delta _{(\vartheta ,\phi )}Y(m_{k};\vartheta ,\phi
)=-b_{l}Y(m_{k};\vartheta ,\phi ),  \label{Spheq}
\end{equation}%
with%
\begin{equation}
b_{l}=l(l+D-2),  \label{bl}
\end{equation}%
from the field equation (\ref{fieldEq}) for the radial function $R_{l}(r)$
one gets%
\begin{equation}
\frac{1}{\sinh ^{D-1}r}\frac{d}{dr}\left( \sinh ^{D-1}r\frac{dR_{l}}{dr}%
\right) +\left[ (E^{2}-m^{2})a^{2}+D(D-1)\xi -\frac{b_{l}}{\sinh ^{2}r}%
\right] R_{l}(r)=0.  \label{Req}
\end{equation}

Introducing a new function $f_{l}(r)=\sinh ^{D/2-1}(r)R_{l}(r)$, we can see
that the general solution for this function is a linear combination of the
associated Legendre functions of the first and second kinds, $%
P_{iz-1/2}^{-\mu }(\cosh r)$ and $Q_{iz-1/2}^{-\mu }(\cosh r)$ (here the
definition of the associated Legendre functions follows that of Ref. \cite%
{Abra72}) with the order and degree determined by%
\begin{equation}
\mu =l+D/2-1,\;z^{2}=E^{2}a^{2}-z_{m}^{2}.  \label{muz}
\end{equation}%
Here and in what follows we use the notation
\begin{equation}
z_{m}=\sqrt{m^{2}a^{2}-D(D-1)\left( \xi -\xi _{D}\right) }.  \label{zm}
\end{equation}%
The relative coefficient in the linear combination depends on the spatial
region under consideration and will be determined below. Now, as a set of
quantum numbers $\alpha $, specifying the mode functions in Eq. (\ref%
{eigfunc1}), we can take $\alpha =$ $(z,l,m_{1},\ldots ,m_{n})$. The energy
is expressed in terms of $z$ by the formula%
\begin{equation}
E(z)=a^{-1}\sqrt{z^{2}+z_{m}^{2}}.  \label{z}
\end{equation}%
Below we shall assume that $z_{m}^{2}\geqslant 0$. In particular, this
condition is satisfied in the most important special cases of minimally and
conformally coupled fields. If it is not obeyed there are modes with
imaginary values of the energy which signal the vacuum instability. Note
that the condition $z_{m}^{2}\geqslant 0$ is different from the
non-negativity condition for the effective mass squared $m^{2}-D(D-1)\xi
/a^{2}$.

The modes (\ref{eigfunc1}) are normalized by the standard orthonormalization
condition
\begin{equation}
\int d^{D}x\,\sqrt{|g|}\varphi _{\alpha }(x)\varphi _{\alpha ^{\prime
}}^{\ast }(x)=\frac{\delta _{\alpha \alpha ^{\prime }}}{2E},
\label{normcond}
\end{equation}%
where the symbol $\delta _{\alpha \alpha ^{\prime }}$ is understood as
Kronecker delta for discrete indices and as the Dirac delta function for
continuous ones. With this normalization, the Wightman function is evaluated
by using the mode-sum%
\begin{equation}
W(x,x^{\prime })=\sum_{\alpha }\varphi _{\alpha }(x)\varphi _{\alpha }^{\ast
}(x^{\prime }),  \label{WFsum}
\end{equation}%
where $\sum_{\alpha }$ stands for the summation over discrete quantum
numbers and for the integration over continuous ones. In what follows, the
Wightman function will be decomposed into the boundary-free and
sphere-induced contributions. For that reason we first consider the
boundary-free geometry.

\subsection{Wightman function in the boundary-free geometry}

For the boundary-free geometry the solution of the radial equation, regular
at the origin, is given in terms of the associated Legendre function of the
first kind. The corresponding positive-energy mode functions have the form
\begin{equation}
\varphi _{\alpha }^{(0)}(x)=A_{\alpha }^{(0)}p_{iz-1/2}^{-\mu
}(u)Y(m_{k};\vartheta ,\phi )e^{-iEt},  \label{phi0}
\end{equation}%
with $0\leqslant z<\infty $ and with the notations
\begin{equation}
p_{\nu }^{-\mu }(u)=\frac{P_{\nu }^{-\mu }(u)}{(u^{2}-1)^{(D-2)/4}}%
,\;u=\cosh r.  \label{pemu}
\end{equation}%
From the property $P_{iz-1/2}^{-\mu }(u)=P_{-iz-1/2}^{-\mu }(u)$ it follows
that the radial function in (\ref{phi0}) is real.

In the case $z=z^{\prime }$ the normalization integral (\ref{normcond}),
with the integration over $r\in \lbrack 0,\infty )$, diverges and, hence,
the main contribution comes from large values $r$. By using the asymptotic
formula for the associated Legendre function for large values of the
argument, we can see that%
\begin{equation}
\int_{1}^{\infty }du\,P_{iz-1/2}^{-\mu }(u)P_{iz^{\prime }-1/2}^{-\mu }(u)=%
\frac{\pi }{z\sinh (\pi z)}\frac{\delta (z-z^{\prime })}{|\Gamma (\mu
+1/2+iz)|^{2}}.  \label{NormInt0}
\end{equation}%
With this result, for the normalization coefficient in (\ref{phi0}) one finds%
\begin{equation}
|A_{\alpha }^{(0)}|^{2}=\frac{z\sinh (\pi z)}{2\pi N(m_{k})a^{D}E}|\Gamma
(\mu +1/2+iz)|^{2}.  \label{Calfa}
\end{equation}%
Here we have used the result $\int d\Omega \,\left\vert Y(m_{k};\vartheta
,\phi )\right\vert ^{2}=N(m_{k})$ (the specific form for $N(m_{k})$ is given
in Ref. \cite{Erde53b} and will not be required in the following
discussion). For $D=3$, the modes (\ref{phi0}) with the coefficient (\ref%
{Calfa}) reduce to the ones discussed in Refs. \cite{Grib94,Grib74}.

Substituting the functions (\ref{phi0}) into the mode-sum (\ref{WFsum}), we
use the addition theorem%
\begin{equation}
\sum_{m_{k}}\frac{Y(m_{k};\vartheta ,\phi )}{N(m_{k})}Y^{\ast
}(m_{k};\vartheta ^{\prime },\phi ^{\prime })=\frac{2l+n}{nS_{D}}%
C_{l}^{n/2}(\cos \theta ),  \label{Addth}
\end{equation}%
where $S_{D}=2\pi ^{D/2}/\Gamma (D/2)$ is the surface area of the unit
sphere in $D$-dimensional space, $C_{l}^{n/2}(\cos \theta )$ is the
Gegenbauer polynomial and $\theta $ is the angle between the directions
determined by $(\vartheta ,\phi )$ and $(\vartheta ^{\prime },\phi ^{\prime
})$. For the corresponding Wightman function we find the formula
\begin{eqnarray}
W_{0}(x,x^{\prime }) &=&\frac{a^{-D}}{2\pi nS_{D}}\sum_{l=0}^{\infty
}(2l+n)C_{l}^{n/2}(\cos \theta )\int_{0}^{\infty }dz\,z\sinh (\pi z)  \notag
\\
&&\times |\Gamma (iz+\mu +1/2)|^{2}p_{iz-1/2}^{-\mu }(u)p_{iz-1/2}^{-\mu
}(u^{\prime })\frac{e^{-iE(z)\Delta t}}{E(z)}.  \label{WF0}
\end{eqnarray}%
with $u^{\prime }=\cosh r^{\prime }$ and with $E(z)$ given by Eq. (\ref{z}).

\subsection{Wightman function inside the sphere}

In the presence of a spherical shell with the radius $r_{0}$, the mode
functions for the interior region, $r<r_{0}$, are written in the form
similar to Eq. (\ref{phi0}):%
\begin{equation}
\varphi _{\alpha }(x)=A_{\alpha }p_{iz-1/2}^{-\mu }(u)Y(m_{k};\vartheta
,\phi )e^{-iE(z)t}.  \label{mode}
\end{equation}%
Imposing the BC (\ref{Robin}), we see that the eigenvalues for the quantum
number $z$ are solutions of the equation%
\begin{equation}
\bar{P}_{iz-1/2}^{-\mu }(u_{0})=0,\;u_{0}=\cosh r_{0}.  \label{Eigeq}
\end{equation}%
Here and in the discussion below, for a given function $F(u)$ we use the
notation%
\begin{equation}
\bar{F}(u)=A(u)F(u)+B(u)F^{\prime }(u),  \label{Barnot}
\end{equation}%
with the coefficients%
\begin{equation}
A(u)=A\sqrt{u^{2}-1}+(D/2-1)\delta _{(j)}\frac{B}{a}u,\;B(u)=-\delta _{(j)}%
\frac{B}{a}(u^{2}-1).  \label{AB}
\end{equation}%
By using the recurrence relation for the associated Legendre function, one
has an equivalent expression%
\begin{eqnarray}
\bar{P}_{iz-1/2}^{-\mu }(u) &=&\left( A\sqrt{u^{2}-1}-\delta _{(j)}l\frac{B}{%
a}u\right) P_{iz-1/2}^{-\mu }(u)  \notag \\
&&+\delta _{(j)}\frac{B}{a}\left[ (1/2+\mu )^{2}+z^{2}\right] \sqrt{u^{2}-1}%
P_{iz-1/2}^{-\mu -1}(u).  \label{Pbar2}
\end{eqnarray}%
For given $r_{0}$ and $l$, the equation (\ref{Eigeq}) has an infinite set of
positive roots with respect to $z$. We shall denote them, arranged in
ascending order of magnitude, as $z=z_{k}$, $k=1,2,\ldots $. Note that these
roots do not depend on the curvature coupling parameter and on the mass of
the field. As it is discussed in appendix \ref{sec:Zeros}, in addition to
the real roots with respect to $z$, depending on the value of $\beta $, a
purely imaginary root may appear. First we consider the case when all the
roots are real. This case is realized for $\beta <\beta _{0}^{(1)}(u_{0})$
(see appendix \ref{sec:Zeros}).

Substituting the mode functions (\ref{mode}) into the orthonormalization
condition, with the integration over the region inside the spherical shell,
for the normalization coefficient one finds%
\begin{equation}
|A_{\alpha }|^{-2}=2Ea^{D}N(m_{k})\int_{1}^{u_{0}}du\,|P_{iz-1/2}^{-\mu
}(u)|^{2}.  \label{C-2}
\end{equation}%
The integral is evaluated by using the integration formula
\begin{equation}
\int_{1}^{b}du\,[P_{iz-1/2}^{-\mu }(u)]^{2}=\frac{b^{2}-1}{2z}\{[\partial
_{z}P_{iz-1/2}^{-\mu }(b)]\partial _{b}P_{iz-1/2}^{-\mu
}(b)-P_{iz-1/2}^{-\mu }(b)\partial _{z}\partial _{b}P_{iz-1/2}^{-\mu }(b)\}.
\label{Intform}
\end{equation}%
For the roots of Eq. (\ref{Eigeq}), $z=z_{k}$, and for $b=u_{0}$, the
expression in the figure braces is equal to $-P_{iz-1/2}^{-\mu
}(u_{0})\partial _{z}\bar{P}_{iz-1/2}^{-\mu }(u_{0})/B(u_{0})$. With this
result, the normalization coefficient is written in terms of $T_{\mu }(z,u)$%
, defined by Eq. (\ref{T2}) in appendix \ref{sec:SF}, by the expression%
\begin{equation}
|A_{\alpha }|^{2}=\frac{e^{i\mu \pi }zT_{\mu }(z,u_{0})}{\pi
a^{D}N(m_{k})E(z)}\Gamma (\mu +iz+1/2)\Gamma (\mu -iz+1/2),  \label{Cnorm2}
\end{equation}%
with $z=z_{k}$.

Having determined the normalized mode functions, we turn to the evaluation
of the Wightman function in the region inside the sphere with the help of
Eq. (\ref{WFsum}). Substituting the eigenfunctions, one finds%
\begin{eqnarray}
W(x,x^{\prime }) &=&\frac{a^{-D}}{\pi nS_{D}}\sum_{l=0}^{\infty
}(2l+n)C_{l}^{n/2}(\cos \theta )e^{i\mu \pi }\sum_{k=1}^{\infty }z_{k}T_{\mu
}(z_{k},u_{0})  \notag \\
&&\times |\Gamma (iz_{k}+\mu +1/2)|^{2}p_{iz_{k}-1/2}^{-\mu
}(u)p_{iz_{k}-1/2}^{-\mu }(u^{\prime })\frac{e^{-iE(z_{k})\Delta t}}{E(z_{k})%
},  \label{WF1}
\end{eqnarray}%
where $u^{\prime }=\sinh r^{\prime }$ and $\Delta t=t-t^{\prime }$. The
roots $z_{k}$ are given implicitly and the representation (\ref{WF1}) is not
convenient for the evaluation of the VEVs of the field squared and the
energy-momentum tensor. Additionally, the terms with large $k$ are highly
oscillatory. Both these difficulties are avoided by applying to the series
over $k$ the summation formula (\ref{SumForm}) with the function%
\begin{equation}
h(z)=z\Gamma (\mu +iz+1/2)\Gamma (\mu -iz+1/2)P_{iz-1/2}^{-\mu
}(u)P_{iz-1/2}^{-\mu }(u^{\prime })\frac{e^{-iE(z)\Delta t}}{E(z)}.
\label{hz}
\end{equation}%
The corresponding conditions are obeyed if $r+r^{\prime }+\Delta t/a<2r_{0}$%
. Note that the function (\ref{hz}) has branch points $z=\pm iz_{m}$.

The part in the Wightman function obtained from the first integral in the
right-hand side of Eq. (\ref{SumForm}) coincides with the boundary-free
function $W_{0}(x,x^{\prime })$. In the second integral, the part over the
interval $(0,z_{m})$ vanishes and for the Wightman function we find%
\begin{equation}
W(x,x^{\prime })=W_{0}(x,x^{\prime })+W_{\mathrm{s}}(x,x^{\prime }),
\label{Wdec}
\end{equation}%
where for the sphere-induced part one has%
\begin{eqnarray}
W_{\mathrm{s}}(x,x^{\prime }) &=&-\frac{a^{1-D}}{\pi nS_{D}}%
\sum_{l=0}^{\infty }(2l+n)C_{l}^{n/2}(\cos \theta )e^{-i\mu \pi
}\int_{z_{m}}^{\infty }dz\,z  \notag \\
&&\times \frac{\bar{Q}_{z-1/2}^{\mu }(u_{0})}{\bar{P}_{z-1/2}^{-\mu }(u_{0})}%
p_{z-1/2}^{-\mu }(u)p_{z-1/2}^{-\mu }(u^{\prime })\frac{\cosh (\sqrt{%
z^{2}-z_{m}^{2}}\Delta t/a)}{\sqrt{z^{2}-z_{m}^{2}}}.  \label{WF2}
\end{eqnarray}%
In deriving this formula we have used the relation \cite{Abra72}%
\begin{equation}
Q_{z-1/2}^{-\mu }(u)=e^{-2i\mu \pi }\frac{\Gamma (z-\mu +1/2)}{\Gamma (z+\mu
+1/2)}Q_{x-1/2}^{\mu }(u),  \label{RelQ}
\end{equation}%
for the associated Legendre function. Formula (\ref{WF2}) provides our final
expression for the sphere-induced part of the Wightman function in the
interior region. In this form the knowledge of the roots $z_{k}$ is not
required and for $r+r^{\prime }+\Delta t/a<2r_{0}$ the integrand
exponentially decays in the upper limit. In the special case $D=3$ and for
Dirichlet BC, by taking into account Eq. (\ref{RelQ}), we see that Eq. (\ref%
{WF2}) is reduced to the expression given in Ref. \cite{Saha08}.

\begin{figure}[tbph]
\begin{center}
\begin{tabular}{cc}
\epsfig{figure=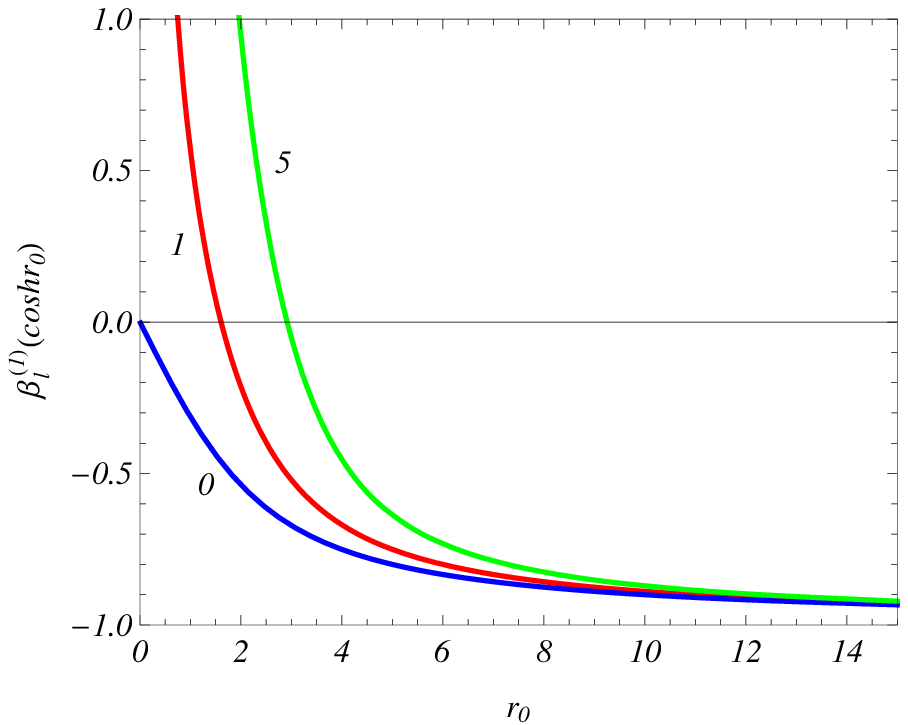,width=7.cm,height=5.5cm} & \quad %
\epsfig{figure=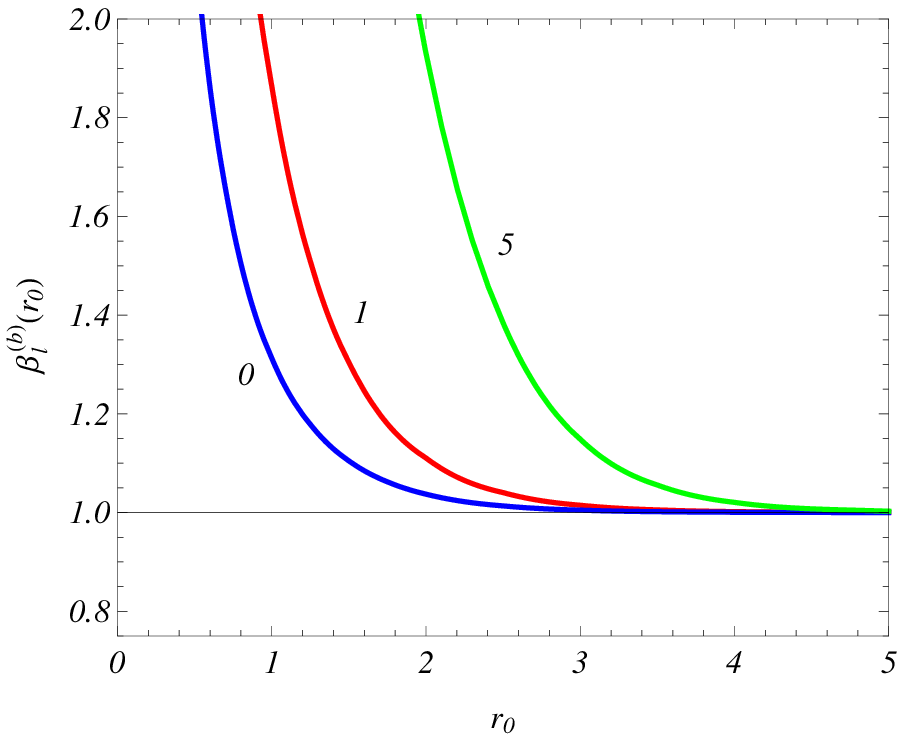,width=7.cm,height=5.5cm}%
\end{tabular}%
\end{center}
\caption{The critical values of the Robin coefficient for the appearance of
the purely imaginary roots of the eigenvalue equation (\protect\ref{Eigeq})
in the interior region (left panel) and for the appearance of the bound
states in the exterior region (right panel) versus the radius of the sphere.
The graphs are plotted for the spatial dimension $D=3$.}
\label{fig1}
\end{figure}

We have considered the case when all the zeros $z_{k}$ of the function $\bar{%
P}_{iz-1/2}^{-\mu }(u_{0})$ are real. As it is noticed in appendix \ref%
{sec:Zeros}, for given $r_{0}$ and $l$, started from some critical value of $%
\beta $, for $\beta >\beta _{l}^{(1)}(\cosh r_{0})$, a single purely
imaginary zero $z=i\eta _{l}$, $\eta _{l}>0$, appears. This zero first
appears for the angular mode $l=0$ and, hence, for a given $r_{0}$ the
purely imaginary zeros are absent if $\beta <\beta _{0}^{(1)}(\cosh r_{0})$.
In order to have a stable vacuum state we assume that $E(i\eta _{l})>0$ or $%
\eta _{l}<z_{m}$. In the left panel of figure \ref{fig1}, for the spatial
dimension $D=3$, we have plotted the critical value $\beta _{l}^{(1)}(\cosh
r_{0})$ for the Robin coefficient, at which the purely imaginary zero
appears, as a function of the sphere radius $r_{0}$ for $l=0,1,5$ (the
numbers near the curves).

In the presence of the imaginary eigenmode, the corresponding contribution
in the Wightman function should be added to the expression (\ref{WF1}). This
contribution is evaluated in a way similar to that we have used for the
modes with real $z$ and has the form%
\begin{equation}
W^{\mathrm{(im)}}(x,x^{\prime })=\sum_{l=0}^{\infty }\frac{2l+n}{nS_{D}a^{D}}%
\frac{C_{l}^{n/2}(\cos \theta )}{u_{0}^{2}-1}\frac{\eta _{l}B(u_{0})}{%
E(i\eta _{l})}\frac{p_{\eta _{l}-1/2}^{-\mu }(u)p_{\eta _{l}-1/2}^{-\mu
}(u^{\prime })e^{-iE(i\eta _{l})\Delta t}}{P_{\eta _{l}-1/2}^{-\mu
}(u_{0})\partial _{\eta }\bar{P}_{\eta -1/2}^{-\mu }(u_{0})|_{\eta =\eta
_{l}}}.  \label{WFim}
\end{equation}%
In the part of the Wightman function corresponding to the contribution of
the modes with real $z$, given by Eq. (\ref{WF1}), again we apply the
summation formula (\ref{SumForm}). But now, as it is explained in appendix %
\ref{sec:Zeros}, in the right-hand side of this formula we should add the
term (\ref{Imzer2}). By taking into account the expression (\ref{hz}) for
the function $h(z)$, we can see that this term exactly cancels the
contribution (\ref{WFim}). Hence, the expression for the Wightman function
given by Eq. (\ref{WF2}) is also valid in the presence of the purely
imaginary zeros for the function $\bar{P}_{iz-1/2}^{-\mu }(u_{0})$ with $%
\eta _{l}<z_{m}$.

As an additional check of Eq. (\ref{WF2}) let us consider the Minkowskian
limit which corresponds to large values of $a$ with fixed $ar=\rho $, where $%
\rho $ is the Minkowskian radial coordinate. Firstly we introduce in Eq. (%
\ref{WF2}) a new integration variable $y=z/a$ and then use the asymptotic
formulae%
\begin{equation}
P_{\nu }^{-\mu }[\cosh (x/\nu )]\approx \frac{I_{\mu }(x)}{\nu ^{\mu }}%
,\;Q_{\nu }^{\mu }[\cosh (x/\nu )]\approx e^{i\mu \pi }\nu ^{\mu }K_{\mu
}(x),  \label{PQlim}
\end{equation}%
valid for $\nu \gg 1$. By taking also into account that for large $a$ one
has $z_{m}=am$, we can see that from Eq. (\ref{WF2}) the expression for the
Wightman function is obtained inside a Robin sphere in Minkowski spacetime
with the radius $\rho _{0}=ar_{0}$ (see Ref. \cite{Saha01}).

\subsection{Exterior region}

In the region outside the sphere, $r>r_{0}$, the radial part of the mode
functions is a linear combination of the two linear independent solution of
the Legendre equation. As such solutions it is convenient to take the
functions $Q_{iz-1/2}^{-\mu }(\cosh r)$ and $Q_{-iz-1/2}^{-\mu }(\cosh r)$.
The relative coefficient of the linear combination of this functions is
determined from the BC (\ref{Robin}) and the mode functions are written in
the form
\begin{equation}
\varphi _{\alpha }(x)=B_{\alpha }\frac{Z_{iz}^{-\mu }(u)Y(m_{k};\vartheta
,\phi )}{(u^{2}-1)^{(D-2)/4}}e^{-iE(z)t},  \label{Eigfuncext}
\end{equation}%
where, as before, $u=\cosh r$, and
\begin{equation}
Z_{iz}^{-\mu }(u)=\bar{Q}_{iz-1/2}^{-\mu }(u_{0})Q_{-iz-1/2}^{-\mu }(u)-\bar{%
Q}_{-iz-1/2}^{-\mu }(u_{0})Q_{iz-1/2}^{-\mu }(u).  \label{Zmu}
\end{equation}%
The notation with the bar is defined by Eq. (\ref{Barnot}) with the
coefficients given in Eq. (\ref{AB}). In addition to the modes with real $z$%
, depending on the values of the coefficients in Robin BC, the modes can be
present with purely imaginary $z$. These modes correspond to bound states.
First we consider the case when the bound states are absent.

From the normalization condition (\ref{normcond}), for the modes (\ref%
{Eigfuncext}) we have%
\begin{equation}
B_{\alpha }^{2}\int_{u_{0}}^{\infty }duZ_{iz}^{-\mu }(u)[Z_{iz^{\prime
}}^{-\mu }(u)]^{\ast }=\frac{\delta (z-z^{\prime })}{2a^{D}N(m_{k})E(z)}.
\label{normint}
\end{equation}%
The $u$-integral diverges in the upper limit for $z=z^{\prime }$ and, hence,
the main contribution comes from large values $u$. So, we can replace the
associated Legendre functions with the arguments $u$ by their asymptotic
expressions for large values of the argument. To find this asymptotic we use
the expression of the associated Legendre function in terms of the
hypergeometric function,
\begin{eqnarray}
Q_{iz-1/2}^{-\mu }(\cosh r) &=&\sqrt{\pi }\frac{\Gamma (1/2+iz-\mu )}{%
e^{i\mu \pi }\Gamma (1+iz)}\frac{e^{-(iz+1/2)r}}{(1-e^{-2r})^{\mu }}  \notag
\\
&&\times F(1/2-\mu ,1/2+iz-\mu ;1+iz;e^{-2r}).  \label{QFrel}
\end{eqnarray}%
From here, for large values $u$ to the leading order we find
\begin{equation}
Q_{iz-1/2}^{-\mu }(u)\approx \sqrt{\pi }e^{-i\mu \pi }\frac{\Gamma
(1/2+iz-\mu )}{\Gamma (1+iz)}\frac{e^{-iz\ln (2u)}}{\sqrt{2u}}.
\label{Qlarge}
\end{equation}%
Substituting this asymptotic into the integral (\ref{normint}), for the
normalization coefficient one finds%
\begin{equation}
B_{\alpha }^{2}=\frac{V^{-2}(z)}{2\pi ^{2}a^{D}N(m_{k})E},  \label{Aalf}
\end{equation}%
where%
\begin{equation}
V(z)=\left\vert e^{i\mu \pi }\bar{Q}_{iz-1/2}^{-\mu }(u_{0})\frac{\Gamma
(1/2-iz-\mu )}{\Gamma (1-iz)}\right\vert .  \label{Bz}
\end{equation}

By using the mode functions (\ref{Eigfuncext}) with the normalization
coefficient (\ref{Aalf}), from the mode-sum formula (\ref{WFsum}) for the
Wightman function in the exterior region we find%
\begin{equation}
W(x,x^{\prime })=\frac{a^{-D}}{2\pi ^{2}nS_{D}}\sum_{l=0}^{\infty }\frac{%
(2l+n)C_{l}^{n/2}(\cos \theta )}{(\sinh r\sinh r^{\prime })^{D/2-1}}%
\int_{0}^{\infty }dz\frac{e^{-iE\Delta t}}{E(z)V^{2}(z)}Z_{iz}^{-\mu
}(u)[Z_{iz}^{-\mu }(u^{\prime })]^{\ast }.  \label{Wext}
\end{equation}%
Introducing instead of the function $Q_{-iz-1/2}^{-\mu }(u)$ the function $%
P_{iz-1/2}^{-\mu }(u)$ and using the properties of the gamma function, the
Wightman function is also presented in the form%
\begin{eqnarray}
W(x,x^{\prime }) &=&\frac{a^{-D}}{2\pi nS_{D}}\sum_{l=0}^{\infty }\frac{%
(2l+n)C_{l}^{n/2}(\cos \theta )}{(\sinh r\sinh r^{\prime })^{D/2-1}}%
\int_{0}^{\infty }dz\,z\sinh (\pi z)  \notag \\
&&\times |\Gamma (iz+\mu +1/2)|^{2}\frac{Y_{iz-1/2}^{-\mu
}(u)[Y_{iz-1/2}^{-\mu }(u^{\prime })]^{\ast }}{\bar{Q}_{iz-1/2}^{-\mu
}(u_{0})[\bar{Q}_{iz-1/2}^{-\mu }(u_{0})]^{\ast }}\frac{e^{-iE(z)\Delta t}}{%
E(z)},  \label{Wext1}
\end{eqnarray}%
where%
\begin{equation}
Y_{\nu }^{-\mu }(u)=P_{\nu }^{-\mu }(u)\bar{Q}_{\nu }^{-\mu }(u_{0})-\bar{P}%
_{\nu }^{-\mu }(u_{0})Q_{\nu }^{-\mu }(u).  \label{Ymun}
\end{equation}

Here we are interested in the part of the Wightman function induced by the
spherical shell. To obtain this contribution we subtract from the function (%
\ref{Wext1}) the Wightman function for the geometry without boundaries which
is given by Eq. (\ref{WF0}). For the further evaluation of the difference we
use the identity
\begin{eqnarray}
&&\frac{Y_{iz-1/2}^{-\mu }(u)[Y_{iz-1/2}^{-\mu }(u^{\prime })]^{\ast }}{\bar{%
Q}_{iz-1/2}^{-\mu }(u_{0})[\bar{Q}_{iz-1/2}^{-\mu }(u_{0})]^{\ast }}%
-P_{iz-1/2}^{-\mu }(u)P_{iz-1/2}^{-\mu }(u^{\prime })=\frac{e^{i\mu \pi }}{%
\pi i\sinh (\pi z)}  \notag \\
&&\qquad \times \sum_{s=\pm 1}s\cos [\pi (siz-\mu )]\frac{\bar{P}%
_{iz-1/2}^{-\mu }(u_{0})}{\bar{Q}_{siz-1/2}^{-\mu }(u_{0})}Q_{siz-1/2}^{-\mu
}(u)Q_{siz-1/2}^{-\mu }(u^{\prime }).  \label{ident1}
\end{eqnarray}%
Substituting this in the expression for the part of the Wightman function
induced by the sphere, we rotate the integration contour over $z$ by the
angle $\pi /2$ for the term with $s=-1$ and by the angle $-\pi /2$ for the
term with $s=1$. This leads to the following result:
\begin{eqnarray}
W_{\mathrm{s}}(x,x^{\prime }) &=&-\frac{a^{1-D}}{\pi nS_{D}}%
\sum_{l=0}^{\infty }(2l+n)C_{l}^{n/2}(\cos \theta )e^{-i\mu \pi
}\int_{z_{m}}^{\infty }dz\,z  \notag \\
&&\times \frac{\bar{P}_{z-1/2}^{-\mu }(u_{0})}{\bar{Q}_{z-1/2}^{\mu }(u_{0})}%
q_{z-1/2}^{\mu }(u)q_{z-1/2}^{\mu }(u^{\prime })\frac{\cosh (\Delta t\sqrt{%
z^{2}-z_{m}^{2}})}{\sqrt{z^{2}-z_{m}^{2}}},  \label{Wsext}
\end{eqnarray}%
with the notation%
\begin{equation}
q_{\nu }^{\mu }(u)=\frac{Q_{\nu }^{\mu }(u)}{(u^{2}-1)^{(D-2)/4}}.
\label{qmu}
\end{equation}%
Here, once again, we have used Eq. (\ref{RelQ}). Comparing the expression (%
\ref{Wsext}) with Eq. (\ref{WF2}), we see that the sphere-induced parts in
the Wightman function for interior and exterior regions are obtained from
each other by the replacements $P_{z-1/2}^{-\mu }\rightleftarrows
Q_{z-1/2}^{\mu }$.

Now, let us turn to the case when bound states are present. For these states
$z=i\eta $, $\eta >0$, and the mode functions are given by the expression%
\begin{equation}
\varphi _{\alpha }(x)=B_{\mathrm{b}\alpha }q_{\eta -1/2}^{-\mu
}(u)Y(m_{k};\vartheta ,\phi )e^{-iE(i\eta )t},  \label{phiB}
\end{equation}%
with the energy $E(i\eta )=a^{-1}\sqrt{z_{m}^{2}-\eta ^{2}}$. In order to
have a stable vacuum state we assume that $\eta <z_{m}$. From the BC (\ref%
{Robin}) we see that, for a given $r_{0}$, the possible bound states are
solutions of the equation%
\begin{equation}
\bar{Q}_{\eta -1/2}^{-\mu }(u_{0})=0.  \label{BSeq}
\end{equation}%
Note that this equation does not involve the field mass and the curvature
coupling parameter. The numerical analysis shows that there are no bound
states for $\beta \leqslant (D-1)/2$. With fixed $r_{0}$ and $l$, a single
bound state appears started from some critical value $\beta
_{l}^{(b)}(r_{0})>(D-1)/2$. We shall denote the corresponding root by $\eta
=\eta _{(b)l}$. This critical value increases with increasing $l$. In the
right panel of figure \ref{fig1}, for the $D=3$ case, we have plotted $\beta
_{l}^{(b)}(r_{0})$ as a function of the sphere radius $r_{0}$ for several
values of $l$ (numbers near the curves). From the asymptotic expression for
the function $Q_{\eta -1/2}^{-\mu }(u_{0})$ it can be seen that $\beta
_{l}^{(b)}(r_{0})\rightarrow (D-1)/2$ for $r_{0}\rightarrow \infty $.

In order to find the normalization coefficient in Eq. (\ref{phiB}) we use
the integration formula%
\begin{equation}
\int_{u}^{\infty }dx\,[Q_{\eta -1/2}^{-\mu }(x)]^{2}=\frac{u^{2}-1}{2\eta }%
\{[\partial _{\eta }Q_{\eta -1/2}^{-\mu }(u)]\partial _{u}Q_{\eta
-1/2}^{-\mu }(u)-Q_{\eta -1/2}^{-\mu }(u)\partial _{\eta }\partial
_{u}Q_{\eta -1/2}^{-\mu }(u)\}.  \label{IntformQ}
\end{equation}%
This formula is obtained by making use of the differential equation for the
associated Legendre function. For the roots of Eq. (\ref{BSeq}) from Eq. (%
\ref{IntformQ}) we get%
\begin{equation}
\int_{u_{0}}^{\infty }dx\,[Q_{\eta -1/2}^{-\mu }(x)]^{2}=\frac{1-u_{0}^{2}}{%
2\eta B(u_{0})}Q_{\eta -1/2}^{-\mu }(u_{0})\partial _{\eta }\bar{Q}_{\eta
-1/2}^{-\mu }(u_{0})|_{\eta =\eta _{(b)l}}.  \label{IntformQ2}
\end{equation}%
By using this result, the normalization coefficient for the bound states is
presented in the form%
\begin{equation}
|B_{\mathrm{b}\alpha }|^{2}=\frac{a^{-D}\eta B(u_{0})}{N(m_{k})E(\eta )}%
\frac{\left( 1-u_{0}^{2}\right) ^{-1}}{Q_{\eta -1/2}^{-\mu }(u_{0})\partial
_{\eta }\bar{Q}_{\eta -1/2}^{-\mu }(u_{0})}|_{\eta =\eta _{(b)l}}.
\label{B2}
\end{equation}%
With this coefficient, for the contribution of the bound states to the
Wightman function one gets%
\begin{equation}
W^{\mathrm{(bs)}}(x,x^{\prime })=\sum_{l=0}^{\infty }\frac{2l+n}{nS_{D}a^{D}}%
\frac{B(u_{0})C_{l}^{n/2}(\cos \theta )}{E(\eta )\left( 1-u_{0}^{2}\right) }%
\frac{\eta q_{\eta -1/2}^{-\mu }(u)q_{\eta -1/2}^{-\mu }(u^{\prime })}{%
Q_{\eta -1/2}^{-\mu }(u_{0})\partial _{\eta }\bar{Q}_{\eta -1/2}^{-\mu
}(u_{0})}e^{-iE(i\eta )\Delta t}|_{\eta =\eta _{(b)l}}.  \label{Wbs}
\end{equation}

In the presence of the bound states, the contribution to the Wightman
function from the modes with real $z$ is still given by Eq. (\ref{Wext1}).
Again, for this expression we use the identity (\ref{ident1}). The
difference from the previous case arises at the step when one rotates the
integration contour over $z$. Now, the integrands have simple poles on the
imaginary axis corresponding to the zeros of the functions $\bar{Q}%
_{siz-1/2}^{-\mu }(u_{0})$ in the denominator of Eq. (\ref{ident1}).
Rotating the integration contours for the terms with $s=-1$ and $s=1$, we
escape these poles by semicircles of small radius in the right-half plane.
The integrals over these semicircles combine in the residue at the point $%
z=e^{\pi i/2}\eta _{(b)l}$. As a result, for the part in the Wightman
function coming from the modes with real $z$ we get the expression (\ref%
{Wsext}) plus the contribution coming from the residue at the pole $z=e^{\pi
i/2}\eta _{(b)l}$. Now, it can be seen that the latter is exactly canceled
by the contribution from the bound state, given by Eq. (\ref{Wbs}). Hence,
we conclude that the expression (\ref{Wsext}) for the shell-induced Wightman
function in the exterior region is valid in the presence of bound states as
well.

Similarly to the case of the interior region, we can see that in the limit $%
a\rightarrow \infty $ from (\ref{Wsext}) the corresponding expression is
obtained for the sphere in Minkowski spacetime.

\section{VEV of the field squared}

\label{sec:phi2}

Among the most important local characteristics of the vacuum state are the
VEVs of the field squared and the energy-momentum tensor. We start with the
mean field squared. It is obtained from the Wightman function in the
coincidence limit of the arguments. This limit is divergent and a
renormalization procedure is required. In curved backgrounds the structure
of divergences is determined by the local geometry. In the problem under
consideration, for points away from the sphere, the local geometry is the
same as in the boundary-free case and, hence, the divergences in the local
physical characteristics are the same as well. From here it follows that the
renormalization procedure for these characteristics is the same as that in
the boundary-free geometry. In the expressions given above for the both
interior and exterior regions we have explicitly decomposed the Wightman
function into the boundary-free and sphere-induced parts. For points outside
the sphere, the renormalization is needed for the boundary-free VEVs only
and the sphere-induced parts are directly obtained from the corresponding
Wigthman function in the coincidence limit.

For the renormalized mean field squared we get%
\begin{equation}
\langle \varphi ^{2}\rangle =\langle \varphi ^{2}\rangle _{0}+\langle
\varphi ^{2}\rangle _{\mathrm{s}},  \label{phi20s}
\end{equation}%
where $\langle \varphi ^{2}\rangle _{0}$ is the renormalized VEV in the
boundary-free space and the part $\langle \varphi ^{2}\rangle _{\mathrm{s}}$
is induced by the sphere. For the latter, by taking into account that $%
C_{l}^{n/2}(1)=\Gamma (l+n)/[\Gamma (n)l!]$, in the interior region from Eq.
(\ref{WF2}) one finds

\begin{equation}
\langle \varphi ^{2}\rangle _{\mathrm{s}}=-\frac{a^{1-D}}{\pi S_{D}}%
\sum_{l=0}^{\infty }D_{l}e^{-i\mu \pi }\int_{z_{m}}^{\infty }dz\,z\frac{\bar{%
Q}_{z-1/2}^{\mu }(u_{0})}{\bar{P}_{z-1/2}^{-\mu }(u_{0})}\frac{%
[p_{z-1/2}^{-\mu }(u)]^{2}}{\sqrt{z^{2}-z_{m}^{2}}},  \label{phi2new}
\end{equation}%
where
\begin{equation}
D_{l}=2\mu \frac{\Gamma (l+D-2)}{\Gamma (D-1)l!},  \label{Dl}
\end{equation}%
is the degeneracy of the angular mode with given $l$. For the functions in
the integrand, we have numerically checked that $e^{-i\mu \pi }\bar{Q}%
_{z-1/2}^{\mu }(u_{0})>0$ in both special cases of Dirichlet and Neumann
BCs, whereas $\bar{P}_{z-1/2}^{-\mu }(u_{0})>0$ for Dirichlet BC and $\bar{P}%
_{z-1/2}^{-\mu }(u_{0})<0$ for Neumann one. Consequently, the sphere-induced
VEV of the field squared is negative for Dirichlet BC and positive for
Neumann BC.

By using the asymptotic formulas%
\begin{equation}
P_{z-1/2}^{-\mu }(u)\approx \frac{z^{-\mu -1/2}e^{rz}}{\sqrt{2\pi \sinh r}}%
,\;Q_{z-1/2}^{\mu }(u)\approx e^{i\mu \pi }\frac{\pi z^{\mu -1/2}e^{-rz}}{%
\sqrt{2\pi \sinh r}},  \label{PQas}
\end{equation}%
valid for $z\gg 1$, we see that the integrand behaves as $e^{2z(r-r_{0})}/z$%
. Hence, the sphere-induced VEV (\ref{phi2new}) diverges on the boundary. In
order to find the leading term in the asymptotic expansion over the distance
from the sphere, we note that for points close to the boundary the dominant
contribution in Eq. (\ref{phi2new}) comes from large values of $z$ and $l$.
In this case, instead of Eq. (\ref{PQas}), we need to use the uniform
asymptotic expansions for the associated Legendre functions for large values
of both $z$ and $\mu $. The latter can be obtained from Eq. (\ref{PQlim}) by
making use of the uniform asymptotic expansions for the modified Bessel
functions (see, for instance, Ref. \cite{Abra72}). In this way, to the
leading order, we get%
\begin{equation}
\langle \varphi ^{2}\rangle _{\mathrm{s}}\approx -\frac{a^{1-D}\Gamma
((D-1)/2)\delta _{B}}{(4\pi )^{(D+1)/2}(r_{0}-r)^{D-1}},  \label{phi2near}
\end{equation}%
with $\delta _{B}=2\delta _{0B}-1$. As is seen, near the sphere the
boundary-induced VEV has opposite signs for Dirichlet and non-Dirichlet BCs.
By taking into account that $a(r_{0}-r)$ is the proper distance from the
sphere, we see that the leading term coincides with that for the sphere in
Minkowski bulk. Of course, this is natural, because near the sphere the
contribution of the modes with the wavelengths smaller than the curvature
radius dominate and they are relatively insensitive to the background
geometry.

At the sphere center, by taking into account the asymptotic $p_{z-1/2}^{-\mu
}(u)\approx 2^{-\mu }r^{l}/\Gamma (\mu +1)$ for $r\rightarrow 0$, we see
that the $l=0$ mode contributes only:%
\begin{equation}
\langle \varphi ^{2}\rangle _{\mathrm{s}}|_{r=0}=\frac{(2a)^{1-D}e^{-i\pi
D/2}}{\pi ^{D/2+1}\Gamma (D/2)}\int_{z_{m}}^{\infty }dz\,\frac{z}{\sqrt{%
z^{2}-z_{m}^{2}}}\frac{\bar{Q}_{z-1/2}^{D/2-1}(u_{0})}{\bar{P}%
_{z-1/2}^{1-D/2}(u_{0})}.  \label{phi2r0}
\end{equation}%
Near the center the contribution of the modes with higher $l$ decays as $%
r^{2l}$. Note that Eq. (\ref{phi2r0}) is further simplified for $D=3$. By
taking into account the formulas%
\begin{equation}
P_{z-1/2}^{-1/2}(u)=\frac{2\sinh (zr)}{z\sqrt{2\pi \sinh r}}%
,\;Q_{z-1/2}^{1/2}(u)=\frac{i\pi e^{-zr}}{\sqrt{2\pi \sinh r}},  \label{PQsp}
\end{equation}%
we obtain%
\begin{equation}
\frac{\bar{Q}_{z-1/2}^{1/2}(u_{0})}{\bar{P}_{z-1/2}^{-1/2}(u_{0})}=\frac{%
i\pi z}{\frac{\beta +u_{0}-z}{\beta +u_{0}+z}e^{2zr_{0}}-1}.  \label{QPrat}
\end{equation}%
Hence, for $D=3$, Eq. (\ref{phi2r0}) is reduced to%
\begin{equation}
\langle \varphi ^{2}\rangle _{\mathrm{s}}|_{r=0}=-\frac{a^{-2}}{2\pi ^{2}}%
\int_{z_{m}}^{\infty }dz\,\frac{z^{2}}{\sqrt{z^{2}-z_{m}^{2}}}\frac{1}{\frac{%
\beta +u_{0}-z}{\beta +u_{0}+z}e^{2zr_{0}}-1},  \label{phi2r0D3}
\end{equation}%
with $z_{m}^{2}=m^{2}a^{2}+1-6\xi $.

In the discussion below we shall also need the covariant Dalambertian of the
boundary-induced part:%
\begin{equation}
\nabla _{p}\nabla ^{p}\langle \varphi ^{2}\rangle _{\mathrm{s}}=\frac{%
2a^{-1-D}}{\pi S_{D}}\sum_{l=0}^{\infty }D_{l}e^{-i\mu \pi
}\int_{z_{m}}^{\infty }dz\,z\frac{\bar{Q}_{z-1/2}^{\mu }(u_{0})}{\bar{P}%
_{z-1/2}^{-\mu }(u_{0})}\frac{F[p_{z-1/2}^{-\mu }(u)]}{\sqrt{z^{2}-z_{m}^{2}}%
}.  \label{Dalambphi2}
\end{equation}%
In this formula we have introduced the function%
\begin{eqnarray}
F[f(u)] &=&\frac{1}{2}\left[ (u^{2}-1)\partial _{u}^{2}+Du\partial _{u}%
\right] f^{2}(u)  \notag \\
&=&(u^{2}-1)f^{\prime 2}(u)+\left[ \frac{b_{l}}{u^{2}-1}-\frac{(D-1)^{2}}{4}%
+z^{2}\right] f^{2}(u).  \label{Ffu}
\end{eqnarray}%
In the second expression, in order to exclude the second order derivative,
we have used the differential equation obeyed by the function $%
p_{z-1/2}^{-\mu }(u)$.

In the exterior region, the shell-induced contribution in the VEV of the
field squared is obtained from Eq. (\ref{Wsext}) in the coincidence limit:%
\begin{equation}
\langle \varphi ^{2}\rangle _{\mathrm{s}}=-\frac{a^{1-D}}{\pi S_{D}}%
\sum_{l=0}^{\infty }D_{l}e^{-i\mu \pi }\int_{z_{m}}^{\infty }dz\,z\frac{\bar{%
P}_{z-1/2}^{-\mu }(u_{0})}{\bar{Q}_{z-1/2}^{\mu }(u_{0})}\frac{%
[q_{z-1/2}^{\mu }(u)]^{2}}{\sqrt{z^{2}-z_{m}^{2}}}.  \label{phi2sext}
\end{equation}%
This quantity is negative for Dirichlet BC and positive for Neumann BC. For
points near the sphere, the leading term in the asymptotic expansion of this
VEV is given by Eq. (\ref{phi2near}) with $r_{0}-r$ replaced by $r-r_{0}$.
In this limit, the effects of the background gravitational field are small.
The curvature effects are crucial at distances from the sphere larger than
the curvature radius of the background space. This corresponds to the limit
of large $r$ with a fixed value of the sphere radius $r_{0}$. For $r\gg 1$
we use the approximate formula
\begin{equation}
q_{z-1/2}^{\mu }(u)\approx 2^{D/2-1}\frac{e^{i\mu \pi }\sqrt{\pi }\Gamma
(z+1/2+\mu )}{\Gamma (z+1)e^{(z+(D-1)/2)r}}.  \label{qmularg}
\end{equation}%
With this asymptotic, the dominant contribution in the integral of Eq. (\ref%
{phi2sext}) comes from the region near the lower limit of the integration.
For $z_{m}>0$, assuming that $z_{m}r\gg 1$, to the leading order we get%
\begin{equation}
\langle \varphi ^{2}\rangle _{\mathrm{s}}\approx -\frac{2^{D-3}a^{1-D}\sqrt{%
\pi z_{m}/r}}{S_{D}e^{(2z_{m}+D-1)r}}\sum_{l=0}^{\infty }D_{l}e^{i\mu \pi }%
\frac{\bar{P}_{z_{m}-1/2}^{-\mu }(u_{0})}{\bar{Q}_{z_{m}-1/2}^{\mu }(u_{0})}%
\frac{\Gamma ^{2}(z_{m}+l+D/2)}{\Gamma ^{2}(z_{m}+1)}.  \label{phi2large}
\end{equation}%
The boundary-induced VEV is exponentially small and the suppression factor
depends on the curvature coupling parameter. For $z_{m}=0$ the leading term
in the asymptotic expansion at large distance takes the form
\begin{equation}
\langle \varphi ^{2}\rangle _{\mathrm{s}}\approx -\frac{2^{D-3}a^{1-D}}{%
S_{D}re^{(D-1)r}}\sum_{l=0}^{\infty }D_{l}e^{i\mu \pi }\frac{\bar{P}%
_{-1/2}^{-\mu }(u_{0})}{\bar{Q}_{-1/2}^{\mu }(u_{0})}\Gamma ^{2}(l+D/2).
\label{phi2largez0}
\end{equation}%
In this case the decay is weaker, though, again exponential. In particular,
for both minimally and conformally coupled massless scalars the suppression
of the boundary induced VEVs at large distances is exponential.

For a spherical boundary in Minkowski bulk and for a massive field the VEV
at large distances is suppressed by the factor $e^{-2m\rho }$ with $\rho $
being the Minkowskian radial coordinate. In this case the suppression factor
does not depend on the curvature coupling. For a massless field in Minkowski
spacetime, the dominant contribution at large distances comes from the
angular mode $l=0$ and the decay of the VEV is of power-law, like $1/\rho
^{2D-3}$ for $D\geqslant 3$. This shows that the suppression of the vacuum
fluctuations in the gravitational field corresponding to the negative
curvature space is stronger compared with the case of the Minkowskian bulk.
A similar feature is observed in the geometry of planar boundaries on
anti-de Sitter bulk (see Ref. \cite{Gold00}), then yielding another example
of a negative curvature space. For de Sitter geometry, having a positive
curvature, the situation is essentially different: the boundary-induced
contributions in the local VEVs decay at large distances as a power-law for
both massive and massless fields.

\section{Energy-momentum tensor}

\label{sec:EMT}

For the evaluation of the VEV of the energy-momentum tensor we use the
formula
\begin{equation}
\langle T_{ik}\rangle =\lim_{x^{\prime }\rightarrow x}\partial _{i^{\prime
}}\partial _{k}W(x,x^{\prime })+\left[ \left( \xi -1/4\right) g_{ik}\nabla
_{p}\nabla ^{p}-\xi \nabla _{i}\nabla _{k}-\xi \mathcal{R}_{ik}\right]
\langle \varphi ^{2}\rangle ,  \label{EMT}
\end{equation}%
where $\mathcal{R}_{ik}$ is the Ricci tensor. For the geometry under
consideration one has (no summation over $p$) $\mathcal{R}%
_{p}^{p}=-(D-1)/a^{2}$ for $p=1,\ldots ,D$, and the remaining components
vanish. In the right-hand side of Eq. (\ref{EMT}) we have used the
expression for the energy-momentum tensor for a scalar field which differs
from the standard one by the term which vanishes on the solutions of the
field equation and does not contribute to the boundary-induced VEV (see Ref.
\cite{Saha04}).

By taking into account the expressions for the Wightman function and for the
VEV of the field squared from the previous section, the vacuum
energy-momentum tensor is presented in the decomposed form%
\begin{equation}
\langle T_{ik}\rangle =\langle T_{ik}\rangle _{0}+\langle T_{ik}\rangle _{%
\mathrm{s}},  \label{Tik0s}
\end{equation}%
where the first and second terms in the right-hand side correspond to the
boundary-free and sphere-induced contributions. Again, for points away the
sphere, the renormalization is needed for the first term only.

In the interior region, after straightforward calculations, we find that the
VEV of the energy-momentum tensor is diagonal with the components (no
summation over $k$)%
\begin{equation}
\langle T_{k}^{k}\rangle _{\mathrm{s}}=\frac{a^{-1-D}}{\pi S_{D}}%
\sum_{l=0}^{\infty }D_{l}e^{-i\mu \pi }\int_{z_{m}}^{\infty }dz\,z\frac{\bar{%
Q}_{z-1/2}^{\mu }(u_{0})}{\bar{P}_{z-1/2}^{-\mu }(u_{0})}\frac{%
F^{(k)}[p_{z-1/2}^{-\mu }(u)]}{\sqrt{z^{2}-z_{m}^{2}}},  \label{Tkk}
\end{equation}%
where we have introduced the notations%
\begin{eqnarray}
F^{(0)}[f(u)] &=&2\left( \xi -1/4\right) F[f(u)]+(z^{2}-z_{m}^{2})f^{2}(u),
\notag \\
F^{(1)}[f(u)] &=&\frac{1}{2}F[f(u)]+2(D-1)\xi uf(u)f^{\prime }(u)  \notag \\
&&-\left[ (D-1)\xi +\frac{b_{l}}{u^{2}-1}-\frac{(D-1)^{2}}{4}+z^{2}\right]
f^{2}(u),  \label{F1} \\
F^{(k)}[f(u)] &=&2(\xi -1/4)F[f(u)]-2\xi uf(u)f^{\prime }(u)  \notag \\
&&+\frac{1}{D-1}\left[ \frac{b_{l}}{u^{2}-1}-(D-1)^{2}\xi \right] f^{2}(u),
\notag
\end{eqnarray}%
with $F[f(u)]$ defined in Eq. (\ref{Ffu}) and $k=2,3,\ldots ,D$ in the last
expression.

We can check that the part in the VEV of the energy-momentum tensor induced
by the spherical shell satisfies the covariant conservation equation $\nabla
_{k}\langle T_{i}^{k}\rangle _{\mathrm{s}}=0$, which for the geometry under
consideration takes the form

\begin{equation}
(u^{2}-1)\frac{\partial }{\partial u}\langle T_{1}^{1}\rangle _{\mathrm{s}%
}+(D-1)u\left( \langle T_{1}^{1}\rangle _{\mathrm{s}}-\langle
T_{2}^{2}\rangle _{\mathrm{s}}\right) =0.  \label{Conteq}
\end{equation}%
In addition, it can be seen that the boundary-induced parts in the VEVs
satisfy the trace relation%
\begin{equation}
\langle T_{k}^{k}\rangle _{\mathrm{s}}=D(\xi -\xi _{D})\nabla _{p}\nabla
^{p}\langle \varphi ^{2}\rangle _{\mathrm{s}}+m^{2}\langle \varphi
^{2}\rangle _{\mathrm{s}}.  \label{tracerel}
\end{equation}%
In particular, for a conformally coupled massless scalar field the
boundary-induced part in the VEV of the energy-momentum tensor is traceless.
The trace anomalies are contained in the boundary-free part.

The sphere-induced contribution in the VEV of the energy-momentum tensor
diverges on the boundary. The leading terms in the expansion over the
distance from the sphere for the energy density and parallel stresses are
found in a way similar to that we have used for the field squared. They are
given by (no summation over $k$)
\begin{equation}
\langle T_{k}^{k}\rangle _{\mathrm{s}}\approx \frac{D\Gamma ((D+1)/2)(\xi
-\xi _{D})}{2^{D}\pi ^{(D+1)/2}[a(r_{0}-r)]^{D+1}}\delta _{B},
\label{Tkknear}
\end{equation}%
for the components $k=0,2,\ldots $. Again, these leading terms coincide with
those for a sphere in the Minkowski bulk. The leading term for the radial
stress vanishes and the next to the leading terms in the relations (\ref%
{PQlim}) should be kept. Easier way is to use the continuity equation (\ref%
{Conteq}) with the result%
\begin{equation}
\langle T_{1}^{1}\rangle _{\mathrm{s}}\approx \left( 1-\frac{1}{D}\right)
\frac{u_{0}(r_{0}-r)}{\sqrt{u_{0}^{2}-1}}\langle T_{0}^{0}\rangle _{\mathrm{s%
}},  \label{T11near}
\end{equation}%
with $\langle T_{0}^{0}\rangle _{\mathrm{s}}$ from Eq. (\ref{Tkknear}).

At the sphere center the terms with $l=0$ and $l=1$ contribute only and one
has (no summation over $k$)%
\begin{equation}
\langle T_{k}^{k}\rangle _{\mathrm{s}}=\frac{a^{-1-D}}{\pi S_{D}}%
\int_{z_{m}}^{\infty }dz\,\frac{z}{\sqrt{z^{2}-z_{m}^{2}}}\sum_{l=0}^{1}%
\frac{D_{l}e^{-i\mu \pi }F_{l}^{(k)}}{2^{2\mu }\Gamma ^{2}(\mu +1)}\frac{%
\bar{Q}_{z-1/2}^{\mu }(u_{0})}{\bar{P}_{z-1/2}^{-\mu }(u_{0})},
\label{EMTcent}
\end{equation}%
where%
\begin{eqnarray}
F_{0}^{(0)} &=&2\left( \xi -\frac{1}{4}\right) \left[ z^{2}-\frac{(D-1)^{2}}{%
4}\right] +z^{2}-z_{m}^{2},  \notag \\
F_{0}^{(k)} &=&\left[ 2\left( 1-\frac{1}{D}\right) \xi -\frac{1}{2}\right] %
\left[ z^{2}-\frac{(D-1)^{2}}{4}\right] -(D-1)\xi ,  \label{F0k}
\end{eqnarray}%
and%
\begin{eqnarray}
F_{1}^{(0)} &=&2D\left( \xi -1/4\right) ,  \notag \\
F_{1}^{(k)} &=&2(D-1)\xi -D/2+1.  \label{F1k}
\end{eqnarray}%
In Eqs. (\ref{F0k}) and (\ref{F1k}), $k=1,2,\ldots ,D$ and the stresses are
isotropic at the center. For $D=3$, Eq. (\ref{EMTcent}) is further
simplified by using Eq. (\ref{QPrat}). The functions $P_{z-1/2}^{-3/2}(u)$
and $Q_{z-1/2}^{3/2}(u)$ in the $l=1$ term of Eq. (\ref{EMTcent}) are
expressed in terms of the hyperbolic functions by using Eq. (\ref{PQsp}) and
the recurrence relations for the associated Legendre functions.

For the shell-induced contribution in the VEV of the energy-momentum tensor
in the exterior region, $r>r_{0}$, we get%
\begin{equation}
\langle T_{k}^{k}\rangle _{\mathrm{s}}=\frac{a^{-1-D}}{\pi S_{D}}%
\sum_{l=0}^{\infty }D_{l}e^{-i\mu \pi }\int_{z_{m}}^{\infty }dz\,z\frac{\bar{%
P}_{z-1/2}^{-\mu }(u_{0})}{\bar{Q}_{z-1/2}^{\mu }(u_{0})}\frac{%
F^{(k)}[q_{z-1/2}^{\mu }(u)]}{\sqrt{z^{2}-z_{m}^{2}}}.  \label{Tkkext}
\end{equation}%
For points near the sphere the corresponding asymptotic for the energy
density and parallel stresses are given by Eq. (\ref{Tkknear}) with the
replacement $(r_{0}-r)\rightarrow (r-r_{0})$. Hence, for a non-conformally
coupled field, near the boundary these components have the same sign in the
exterior and interior regions. For the radial stress we have the same
relation (\ref{T11near}) and for a non-conformally coupled field it has
opposite signs inside and outside the sphere.

Now we turn to the asymptotic behavior of the vacuum energy-momentum tensor
at large distances from the sphere. In this limit we use Eq. (\ref{qmularg}%
). The dominant contribution to the integral in Eq. (\ref{Tkkext}) comes
form the region near the lower limit. For $z_{m}>0$, to the leading order we
get (no summation over $k$)%
\begin{eqnarray}
\langle T_{k}^{k}\rangle _{\mathrm{s}} &=&\frac{2^{D-3}\sqrt{\pi z_{m}/r}%
F^{(k)}(z_{m})}{S_{D}a^{D+1}e^{(2z_{m}+D-1)r}}\sum_{l=0}^{\infty
}D_{l}e^{i\mu \pi }  \notag \\
&&\times \frac{\bar{P}_{z_{m}-1/2}^{-\mu }(u_{0})}{\bar{Q}_{z_{m}-1/2}^{\mu
}(u_{0})}\frac{\Gamma ^{2}(z_{m}+1/2+\mu )}{\Gamma ^{2}(z_{m}+1)},
\label{TkkLarg}
\end{eqnarray}%
where we have defined the functions%
\begin{eqnarray}
F^{(0)}(z) &=&\left( 4\xi -1\right) z[z+(D-1)/2],  \notag \\
F^{(1)}(z) &=&-\frac{1}{2}(D-1)\left[ \left( 4\xi -1\right) z+2D\left( \xi
-\xi _{D}\right) \right] ,  \label{F1z}
\end{eqnarray}%
and $F^{(k)}(z)=2zF^{(1)}(z)/(1-D)$ for $k=2,\ldots ,D$. For minimally and
conformally coupled fields $F^{(0)}(z)<0$ and from Eq. (\ref{TkkLarg}) it
follows that the sphere-induced contribution to the vacuum energy of these
fields is negative for Dirichlet BC and positive for Neumann BC.

For $z_{m}=0$, the leading term in the asymptotic expansion at large
distances has the form (no summation over $k$)%
\begin{equation}
\langle T_{k}^{k}\rangle _{\mathrm{s}}=\frac{2^{D-3}DF_{\mathrm{(e)}%
}^{(k)}e^{-\left( D-1\right) r}}{S_{D}a^{D+1}r^{2-\delta _{1}^{k}}}%
\sum_{l=0}^{\infty }D_{l}e^{i\mu \pi }\frac{\bar{Q}_{-1/2}^{\mu }(u_{0})}{%
\bar{P}_{-1/2}^{-\mu }(u_{0})}\Gamma ^{2}(1/2+\mu ),  \label{TkkLarg2}
\end{equation}%
with the coefficients%
\begin{eqnarray}
F_{\mathrm{(e)}}^{(0)} &=&\xi _{D}\left( 4\xi -1\right) ,  \notag \\
F_{\mathrm{(e)}}^{(1)} &=&(1-D)\left( \xi -\xi _{D}\right) ,\;F_{\mathrm{(e)}%
}^{(k)}=\xi -\xi _{D},  \label{F1e}
\end{eqnarray}%
and with $k=2,\ldots ,D$ in the last expression. In this case, for
non-conformally coupled fields one has $|\langle T_{1}^{1}\rangle _{\mathrm{s%
}}|\gg |\langle T_{0}^{0}\rangle _{\mathrm{s}}|$. Again, at large distances
we have an exponential suppression for both massive and massless fields. For
minimally and conformally coupled fields the energy density corresponding to
Eq. (\ref{TkkLarg2}) is negative for Dirichlet BC and positive for Neumann
one. Note that for a sphere in Minkowski bulk the Casimir densities for
massive fields are suppressed by the factor $e^{-2m\rho }$. For
non-conformally coupled massless fields, at large distances the contribution
of the $l=0$ mode dominates and all the diagonal components of the vacuum
energy-momentum tensor in Minkowski bulk are of the same order. In this case
the VEVs decay as $1/\rho ^{2D-1}$ for $D\geqslant 3$ and as $1/(\rho
^{3}\ln \rho )$ for $D=2$ (see Ref. \cite{Saha01}).

\begin{figure}[tbph]
\begin{center}
\begin{tabular}{cc}
\epsfig{figure=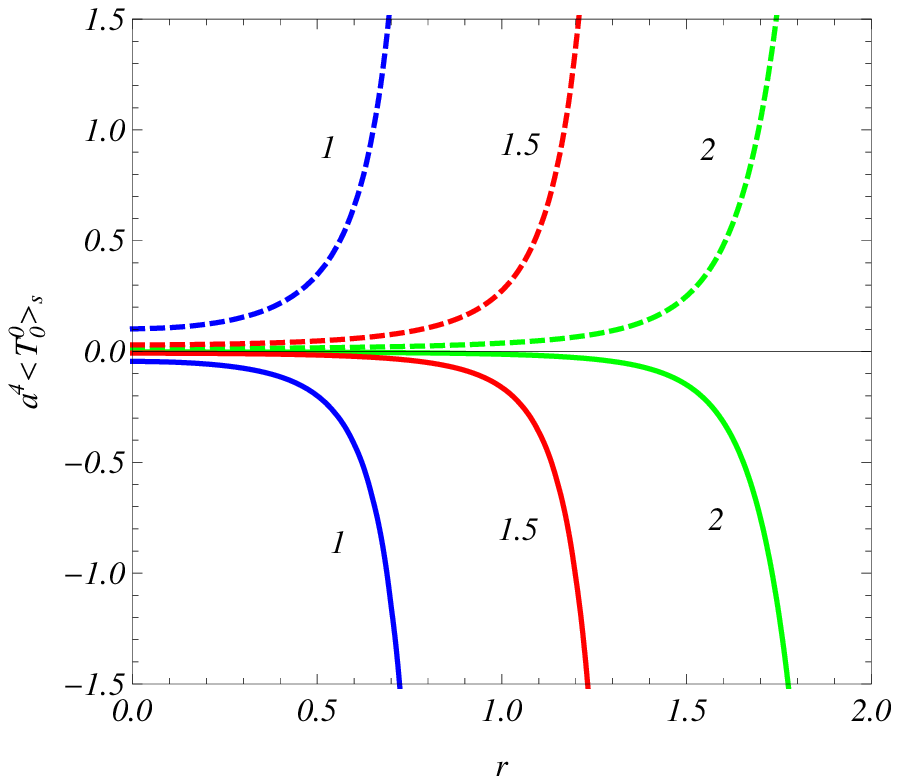,width=7.cm,height=5.5cm} & \quad %
\epsfig{figure=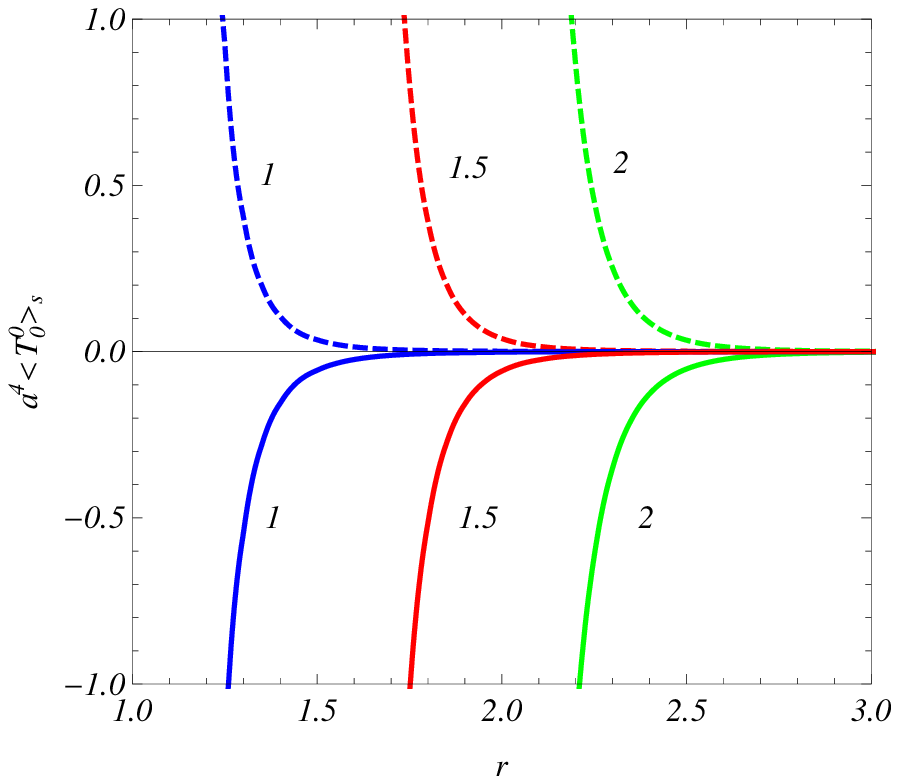,width=7.cm,height=5.5cm}%
\end{tabular}%
\end{center}
\caption{The sphere-induced vacuum energy density for a minimally coupled $%
D=3$ massless scalar field as a function of the radial coordinate. The
graphs are plotted for several values of the sphere radius ($r_{0}=1,1.5,2$,
numbers near the curves). The left/right panel corresponds to the
interior/exterior regions and the full and dashed curves are for Dirichlet
and Neumann BCs, respectively. }
\label{fig2}
\end{figure}

In figure \ref{fig2}, for a minimally coupled massless field, we have
plotted the sphere-induced contribution in the vacuum energy density inside
(left panel) and outside (right panel) the $D=3$ spherical shell versus the
radial coordinate $r$ for the values of the radius $r_{0}=1,1.5,2$ (numbers
near the curves). The full/dashed curves correspond to Dirichlet/Neumann BCs.

\begin{figure}[tbph]
\begin{center}
\epsfig{figure=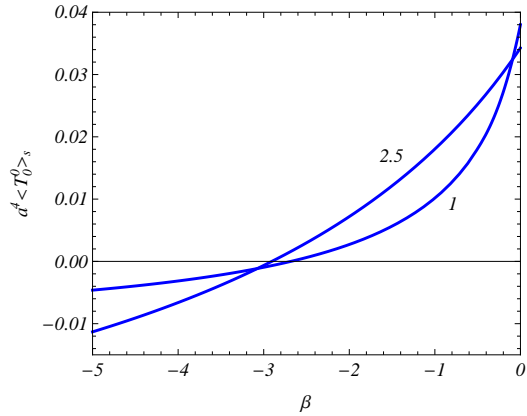,width=7.cm,height=5.5cm}
\end{center}
\caption{The sphere-induced vacuum energy density versus the coefficient in
the Robin BC in the case of a minimally coupled $D=3$ massless field. For
the sphere radius one has $r_{0}=2$ and the numbers near the curves
correspond to the values of the radial coordinate $r$.}
\label{fig3}
\end{figure}

In figure \ref{fig3}, the sphere-induced contribution in the vacuum energy
is displayed as a function of the parameter $\beta $ in Robin BC for fixed
value of $r$ (numbers near the curves) and for $D=3$ sphere with the radius $%
r_{0}=2$. Again, the graphs are plotted for a minimally coupled massless
field.

Here we have considered the VEV of the bulk energy-momentum tensor. On
manifolds with boundaries and for a scalar field with Robin BC there is also
a surface energy-momentum tensor located on the boundary. In the general
case of bulk and boundary geometries the expression for the latter is given
in Ref. \cite{Saha04} (see also the discussion in Ref. \cite{Kenn80}). The
VEV of the surface energy-momentum tensor can be evaluated by making use of
the corresponding mode-sum with the eigenfunctions given above. However, in
this case the renormalization procedure is not reduced to the one in the
boundary-free space and additional subtractions are necessary. This
procedure can be realized, for example, by using the generalized zeta
function technique and will be discussed elsewhere with the total Casimir
energy.

For a conformally coupled massless scalar field, the generalization of the
the results given above for the Friedmann-Robertson-Walker backgrounds,
described by Eq. (\ref{ds2}) with a time-dependent scale factor $a=a(t)$, is
straightforward. For that, in the expressions for the Wightman function and
for the VEVs of the field squared and the energy-momentum tensor one should
make the replacement $a\rightarrow a(t)$. For non-conformally coupled fields
the problem is more complicated. In particular, similar to the case of de
Sitter background, we expect that in addition to the diagonal components,
the vacuum energy-momentum tensor will have an off-diagonal component
describing the energy flux along the radial direction.

\section{Casimir densities induced by spherical bubbles}

\label{sec:Bubble}

In this section we consider the background spacetime with the geometry
described by distinct metric tensors inside and outside a spherical
boundary. The interior and exterior gravitational backgrounds may correspond
to different vacuum states of a physical system. In this case the sphere
serves as a thin-wall approximation of a domain wall interpolating between
two coexisting vacua \cite{Klin05}. In what follows we shall refer the
interior region as a bubble.

\subsection{Bubble in a constant curvature space}

First we consider the geometry described by the line element (\ref{ds2}) in
the region $r>r_{0}$ and by
\begin{equation}
ds^{2}=e^{2u(r)}dt^{2}-a^{2}[e^{2v(r)}dr^{2}+e^{2w(r)}d\Omega _{D-1}^{2}],
\label{ds2in}
\end{equation}%
in the region $r<r_{0}$, assuming that the corresponding metric tensor is
regular. For generality, the presence of the surface energy-momentum tensor
with nonzero components $\tau _{0}^{0}$ and $\tau _{2}^{2}=\cdots =\tau
_{D}^{D}$, located at $r=r_{0}$, will be assumed. The continuity of the
metric tensor at the separating boundary gives%
\begin{equation}
u(r_{0})=0,\;v(r_{0})=0,\;w(r_{0})=\ln (\sinh r_{0}).  \label{metricCont}
\end{equation}%
From the Israel matching conditions on the sphere $r=r_{0}$ we get (no
summation over $k=2,3,\ldots ,D$)%
\begin{eqnarray}
8\pi Ga\tau _{0}^{0} &=&\left( D-1\right) \left[ w^{\prime }(r_{0})-\coth
r_{0}\right] ,  \notag \\
8\pi Ga\tau _{k}^{k} &=&u^{\prime }(r_{0})+\left( D-2\right) \left[
w^{\prime }(r_{0})-\coth r_{0}\right] .  \label{surfemt}
\end{eqnarray}%
where $G$ is the gravitational constant. For the trace of the surface
energy-momentum tensor one has%
\begin{equation}
\frac{8\pi G}{D-1}a\tau =u^{\prime }(r_{0})+(D-1)[w^{\prime }(r_{0})-\coth
r_{0}].  \label{Tracesurf}
\end{equation}%
A general problem for the Casimir densities in spherically symmetric spaces
with bubbles is considered in Ref. \cite{Bell14}. In that paper the exterior
metric was assumed to be asymptotically flat. The latter is not the case for
the problem under consideration. However, the main steps for the evaluation
are similar and we omit the details.

The mode functions for a scalar field in the interior and exterior regions
are presented in the form (\ref{eigfunc1}). We denote the regular solution
of the equation for the radial function in the interior region by $%
R_{(i)\,l}(r,E)$ taking it being real. By taking into account that the
energy appears in the equation for the radial function in the form $E^{2}$,
we shall also assume that $R_{(i)\,l}(r,-E)=\mathrm{const}\cdot
R_{(i)\,l}(r,E)$. For the radial function in the exterior region one has%
\begin{equation}
R_{l}(r)=A_{1}q_{iz-1/2}^{-\mu }(u)+A_{2}q_{-i\lambda -1/2}^{-\mu }(u).
\label{Rext}
\end{equation}%
The radial functions are continous at $r=r_{0}$ and for their derivatives
one has the jump condition \cite{Bell14}%
\begin{equation}
R_{l}^{\prime }(r_{0}+0)-R_{l}^{\prime }(r_{0}-0)=\frac{16\pi G\xi }{D-1}%
a\tau R(r_{0}).  \label{DerJump}
\end{equation}%
This jump comes from the delta function term in the field equation (\ref%
{fieldEq}) contained in the Ricci scalar. From these matching conditions,
for the radial functions in the geometry at hand one gets%
\begin{equation}
R_{l}(r)=C_{\alpha }V_{iz}^{-\mu }(u),\;r>r_{0},  \label{Rext2}
\end{equation}%
where%
\begin{equation}
V_{iz}^{-\mu }(u)=\hat{q}_{-iz-1/2}^{-\mu }(u_{0})q_{iz-1/2}^{-\mu }(u)-\hat{%
q}_{iz-1/2}^{-\mu }(u_{0})q_{-iz-1/2}^{-\mu }(u),  \label{V}
\end{equation}%
and
\begin{equation}
C_{\alpha }=\frac{R_{(i)\,l}(r_{0},E)}{W_{l}^{(12)}(r_{0})}.  \label{B}
\end{equation}%
In Eq. (\ref{V}), the notation with hat is defined as%
\begin{equation}
\hat{F}(u)=\sqrt{u^{2}-1}F^{\prime }(u)-\left[ \frac{R_{(i)\,l}^{\prime
}(r_{0},E)}{R_{(i)\,l}(r_{0},E)}+\frac{16\pi G\xi }{D-1}a\tau \right] F(u),
\label{hat}
\end{equation}%
where $R_{(i)\,l}^{\prime }(r_{0},E)=\partial _{r}R_{(i)\,l}(r,E)|_{r=r_{0}}$%
, and
\begin{equation}
W_{l}^{(12)}(r)=\sqrt{u^{2}-1}W\{q_{iz-1/2}^{-\mu }(u),q_{-iz-1/2}^{-\mu
}(u)\},  \label{Wl12}
\end{equation}%
with $W\{f(u),g(u)\}$ being the Wronskian for the enclosed functions. In the
normalization condition, similar to Eq. (\ref{normcond}), the integration
goes over the both interior and exterior regions. However, the dominant
contribution comes from large values of $r$. By using the asymptotic for the
associated Legendre function, similar to the case of the region outside a
spherical boundary with Robin BC, one gets%
\begin{equation}
C_{\alpha }^{2}=\frac{|\hat{q}_{iz-1/2}^{-\mu }(u_{0})|^{-2}|\Gamma
(1-iz)|^{2}}{2\pi ^{2}a^{D}N(m_{k})E(z)|\Gamma (1/2-iz-\mu )|^{2}}.
\label{Balf2}
\end{equation}

Now substituting the exterior mode functions with the normalization
coefficient (\ref{Balf2}) into the mode-sum for the Wightman function, in
the exterior region we get the formula%
\begin{eqnarray}
W(x,x^{\prime }) &=&\sum_{l=0}^{\infty }\frac{(2l+n)C_{l}^{n/2}(\cos \theta )%
}{2\pi nS_{D}a^{D}}\int_{0}^{\infty }dz\,z\sinh (\pi z)  \notag \\
&&\times |\Gamma (iz+\mu +1/2)|^{2}\frac{y_{iz-1/2}^{-\mu
}(u)[y_{iz-1/2}^{-\mu }(u^{\prime })]^{\ast }}{\hat{q}_{iz-1/2}^{-\mu
}(u_{0})[\hat{q}_{iz-1/2}^{-\mu }(u_{0})]^{\ast }}\frac{e^{-iE(z)\Delta t}}{%
E(z)},  \label{Wextbub}
\end{eqnarray}%
with the function
\begin{equation}
y_{iz-1/2}^{-\mu }(u)=\hat{q}_{iz-1/2}^{-\mu }(u_{0})p_{iz-1/2}^{-\mu }(u)-%
\hat{p}_{iz-1/2}^{-\mu }(u))q_{iz-1/2}^{-\mu }(u).  \label{yf}
\end{equation}%
In deriving Eq. (\ref{Wextbub}) we have used the relation%
\begin{equation}
V_{iz}^{-\mu }(u)=-i\pi \frac{e^{-i\mu \pi }\sinh (\pi z)}{\cos [\pi (iz+\mu
)]}y_{iz-1/2}^{-\mu }(u).  \label{RelVy}
\end{equation}%
Now by comparing Eq. (\ref{Wextbub}) with the expression (\ref{Wext1}) for
the corresponding function outside the Robin sphere, we see that Eq. (\ref%
{Wextbub}) differs from Eq. (\ref{Wext1}) by the replacement
\begin{equation}
\beta \rightarrow -\frac{R_{(i)\,l}^{\prime }(r_{0},E)}{R_{(i)\,l}(r_{0},E)}-%
\frac{16\pi G\xi }{D-1}a\tau .  \label{RobRepl}
\end{equation}

The further evaluation of the Wightman function is similar to that we have
given in section \ref{sec:WF} for the exterior region. The only difference
is that, after using the identity (\ref{ident1}) in the integral
representation (\ref{Wext1}), in the rotation of the contours of the
integrations we should take into account that now the effective Robin
coefficient depends on $z$ through $E=E(z)$ (see Eq. (\ref{RobRepl})).
Because of our choice of the interior radial function, the logarithmic
derivative in Eq. (\ref{RobRepl}) is an even function of the energy and the
dependence on $z$ gives no additional difficulties in the evaluation
process. As a result, the bubble-induced part in the Wightman function is
given by Eq. (\ref{Wsext}) where now, in the notation (\ref{Barnot}), we
should make the replacement%
\begin{equation}
\beta \rightarrow -\frac{R_{(i)\,l}^{\prime }(r_{0},e^{\pi i/2}\sqrt{%
z^{2}-z_{m}^{2}})}{R_{(i)\,l}(r_{0},e^{\pi i/2}\sqrt{z^{2}-z_{m}^{2}})}-%
\frac{16\pi G\xi }{D-1}a\tau ,  \label{RobRepl2}
\end{equation}%
where the second term in the right-hand side is given by Eq. (\ref{Tracesurf}%
). With the same replacement in Eqs. (\ref{phi2sext}) and (\ref{Tkkext}), we
obtain the VEVs of the field squared and the energy-momentum tensor.

Similar to the case of the Robin sphere, in the geometry of the bubble under
consideration, the VEVs of the field squared and the energy-momentum tensor
diverge on the separating boundary. It can be seen that (see also Ref. \cite%
{Bell14}), because of the dependence of the effective Robin coefficient in
Eq. (\ref{RobRepl2}) on the integration variable $z$, the leading terms in
the asymptotic expansion of the VEVs over the distance from the boundary
vanish and the divergences are weaker compared to the case of the Robin
sphere. In particular, the VEV of the field squared diverges as $%
(r-r_{0})^{2-D}$ and the energy density behaves as $(r-r_{0})^{-D}$.

As a simple application of the general results, consider a bubble with the
Minkowskian interior with the functions $v(r)=0$ and $w(r)=\ln r$ in Eq. (%
\ref{ds2in}). For the surface energy density one finds%
\begin{equation}
\tau _{0}^{0}=\frac{D-1}{8\pi Ga}\left( 1/r_{0}-\coth r_{0}\right) ,
\label{tauMink}
\end{equation}%
and for the stresses we get (no summation over $k=2,\ldots ,D$) $\tau
_{k}^{k}=(D-2)\tau _{0}^{0}/(D-1)$. Note that $-1<1/r_{0}-\coth
r_{0}\leqslant 0$ for $\infty >r_{0}\geqslant 0$ and the corresponding
energy density is always negative. In this special case for the regular
radial function in the interior region one has%
\begin{equation}
R_{(i)\,l}(r,E)=\frac{\mathrm{const}}{r^{n/2}}J_{l+n/2}(ar\sqrt{E^{2}-m^{2}}%
),  \label{RiMink}
\end{equation}%
with $J_{\nu }(x)$ being the Bessel function. Hence, the Casimir densities
in the region $r>r_{0}$, induced by the interior Minkowskian geometry are
given by Eqs. (\ref{phi2sext}) and (\ref{Tkkext}), making the replacement%
\begin{equation}
\beta \rightarrow -\frac{1}{r_{0}}\left[ y\frac{I_{l+n/2}^{\prime }(y)}{%
I_{l+n/2}(y)}-\frac{n}{2}+2\xi (D-1)\left( 1-r_{0}\coth r_{0}\right) \right]
,  \label{RobReplM}
\end{equation}%
with the notation%
\begin{equation}
y=r_{0}\sqrt{z^{2}-z_{m}^{2}+m^{2}a^{2}},  \label{y}
\end{equation}%
and with the modified Bessel function $I_{\nu }(x)$. For the example of a
minimally coupled massless scalar field we have checked that there are no
bound states in the geometry under consideration.

\subsection{Bubble of a constant curvature in Minkowski spacetime}

Now we consider a background geometry with the line element (\ref{ds2}) in
the region $r<r_{0}$ and with the Minkowskian spacetime in the region $%
r>r_{0}$. The components of the surface energy-momentum tensor are given by
Eq. (\ref{tauMink}) with the opposite signs. In particular, the
corresponding energy density is always positive. The Casimir densities in
the exterior region are obtained from the general results of Ref. \cite%
{Bell14} for the special interior geometry under consideration.

For the Wightman function one has the decomposition%
\begin{equation}
W(x,x^{\prime })=W_{M}(x,x^{\prime })+W_{\mathrm{b}}(x,x^{\prime }),
\label{WbubC}
\end{equation}%
where $W_{M}(x,x^{\prime })$ is the Wightman function in the Minkowski
spacetime and the part%
\begin{eqnarray}
W_{\mathrm{b}}(x,x^{\prime }) &=&-\sum_{l=0}^{\infty }\frac{\left(
2l+n\right) C_{l}^{n/2}(\cos \theta )}{\pi nS_{D}a^{D-1}\left( rr^{\prime
}\right) ^{n/2}}\int_{ma}^{\infty }dz\,z\frac{\tilde{I}_{l+n/2}(zr_{0},z)}{%
\tilde{K}_{l+n/2}(zr_{0},z)}  \notag \\
&&\times \frac{\cosh ((\Delta t/a)\sqrt{z^{2}-m^{2}a^{2}})}{\sqrt{%
z^{2}-m^{2}a^{2}}}K_{l+n/2}(zr)K_{l+n/2}(zr^{\prime }),  \label{WcM}
\end{eqnarray}%
is induced by the constant curvature bubble. In Eq. (\ref{WcM}), we have
defined the notation
\begin{equation}
\tilde{F}(y,z)=y\partial _{y}F(y)-\left[ r_{0}\sinh r_{0}\frac{%
p_{w(z)-1/2}^{-\mu \prime }(u_{0})}{p_{w(z)-1/2}^{-\mu }(u_{0})}+2\xi
(D-1)\left( r_{0}\coth r_{0}-1\right) +\frac{n}{2}\right] F(y),  \label{FtM}
\end{equation}%
with
\begin{equation}
w(z)=\sqrt{z^{2}-m^{2}a^{2}+z_{m}^{2}}.  \label{wx}
\end{equation}%
Note that under the condition $z_{m}^{2}\geqslant 0$ the function $w(z)$ is
real and the Minkowskian vacuum in the exterior region is stable. The
corresponding Casimir densities in the exterior region are obtained from the
expressions in Ref. \cite{Bell14} by the replacements $\tilde{F}%
(y)\rightarrow \tilde{F}(y,z)$ with Eq. (\ref{FtM}) and $%
F=I_{l+n/2},K_{l+n/2}$.

\section{Conclusion}

\label{sec:Conclus}

We have investigated the properties of the scalar vacuum in a constant
negative curvature space induced by a spherical boundary on which the field
operator obeys Robin BC. General values of the spatial dimension and of the
curvature coupling parameter are considered. For the coefficient in the
Robin BC there is a critical value above which the scalar vacuum becomes
unstable. The properties of the quantum vacuum are encoded in two-point
functions. We have considered the positive-frequency Wightman function and
the other function can be evaluated in a similar way. In order to evaluate
the Wightman function we employed the direct summation over the complete set
of modes. In the region inside the sphere, the eigenmodes of the field are
expressed in terms of the zeros of the combination of the associated
Legendre function and its derivative with respect to the order (see Eq. (\ref%
{Eigeq})). In the mode-sum for the Wightman function, for the summation of
the series over these zeros we have used the formula (\ref{SumForm}),
derived from the generalized Abel-Plana formula. This allowed us to separate
the part corresponding to the boundary-free geometry and to present the
sphere-induced part in terms of a rapidly convergent integral. In this form
the explicit knowledge of the eignemodes is not required. In addition, with
the decomposition into the boundary-free and boundary-induced contributions,
the renormalization of the VEVs in the coincidence limit is reduced to the
one for the boundary-free geometry. We have provided a similar decomposition
for the exterior region as well.

As an important local characteristic of the vacuum state, in section \ref%
{sec:phi2} we consider the VEV of the field squared. The corresponding
expressions for the sphere-induced parts in the interior and exterior
regions are given by Eqs. (\ref{phi2new}) and (\ref{phi2sext}). These parts
are negative for Dirichlet BC and positive for Neumann BC. The VEV of the
field squared diverges on the boundary. The leading term in the asymptotic
expansion over the distance from the sphere is given by Eq. (\ref{phi2near})
and coincides with that for a sphere in Minkowski bulk. In this limit the
dominant contribution comes from the wavelengths smaller than the curvature
radius of the background and the effects of the gravitational field are
small. In the opposite limit of large distances, the effects of gravity are
decisive. The asymptotic behavior in this region depends on whether the
parameter $z_{m}$, defined by Eq. (\ref{zm}), is zero or not. For $z_{m}>0$,
the leading term is given by Eq. (\ref{phi2large}) and the boundary-induced
VEV is exponentially suppressed. For $z_{m}=0$ the leading term is given by
Eq. (\ref{phi2largez0}). The decay in this case is again exponential, though
relatively weaker. Consequently, for both massive and massless fields the
suppression of the boundary-induced VEVs at large distances is exponential.
This is in contrast to the case of Minkowskian bulk, where the decay for
massless fields has a power-law behavior.

Another important characteristic of the vacuum is the VEV of the
energy-momentum tensor. This VEV is diagonal and the expressions for the
sphere-induced contributions are given by Eqs. (\ref{Tkk}) and (\ref{Tkkext}%
) for the interior and exterior regions, respectively. Near the sphere, the
leading term in the energy density and parallel stresses is given by Eq. (%
\ref{Tkknear}). Once again, this term coincides with the corresponding
asymptotic for a spherical shell in flat spacetime. For the leading term in
the normal stress one has the expression (\ref{T11near}) and the
corresponding divergence is weaker. The asymptotics at large distances are
given by Eqs. (\ref{TkkLarg}) and (\ref{TkkLarg2}) for the cases $z_{m}>0$
and $z_{m}=0$, respectively. In both cases the VEVs are suppressed by the
factor $e^{-(2z_{m}+D-1)r}$. At large distances and for non-conformally
coupled fields one has $|\langle T_{1}^{1}\rangle _{\mathrm{s}}|\gg |\langle
T_{0}^{0}\rangle _{\mathrm{s}}|$. Again, in contrast to the Minkowskian
case, the VEVs are exponentially suppressed for both massive and massless
fields. This feature is observed also in the Casimir problems on the
background of anti-de Sitter spacetime, which presents another example of a
negative curvature space. In de Sitter spacetime, having a positive
curvature, the situation is opposite: at large distances from the boundary
the local VEVs for both massive and massless fields exhibit a power-law
behavior.

We have generalized the results for the VEVs to the backgrounds described by
two distinct spherically symmetric metric tensors in the regions separated
by a spherical boundary. The geometry of one region affects the properties
of the vacuum in the other region leading to the gravitationally induced
Casimir effect. We have considered the interior geometry described by the
line element (\ref{ds2in}) and the exterior geometry, as before, corresponds
to a constant curvature space. In addition, we have assumed the presence of
the surface energy-momentum tensor located on the separating boundary. The
corresponding jump conditions on the normal derivatives of the metric tensor
are obtained from the Israel matching conditions, whereas the BCs on the
field operator are obtained from the field equation. In this way, we have
shown that the bubble-induced contributions in the Wightman function and in
the VEVs of the field squared and the energy-momentum tensor are obtained
from the expressions for the spherical shell with Robin BC by the
replacement (\ref{RobRepl2}) of the coefficient. As an example of the
exterior geometry we have considered the case of the Minkowskian bubble. In
this case the replacement needed is given by Eq. (\ref{RobReplM}). Note that
in the geometry with bubbles the divergences on the boundary are weaker
compared with the case of the Robin sphere. We have also considered a bubble
with a negative constant curvatures space in the Minkowski bulk. The
corresponding expressions for the local characteristics are obtained from
the general formulas in Ref. \cite{Bell14} with the notation with tilde
defined by Eq. (\ref{FtM}).

The generalization of the results given above for the
Friedmann-Robertson-Walker cosmological models with a time-dependent scale
factor is straightforward in the case of a conformally coupled massless
field. This requires the replacement $a\rightarrow a(t)$ in the expressions
for the Wightman function and for the VEVs. In the special case $D=2$, the
results obtained can be applied to negatively curved graphene structures
(for this type of structures see Ref. \cite{Iori13}), described in the
long-wavelength regime by an effective 3-dimensional relativistic field
theory. The latter, in addition to the well-known Dirac fermions describing
the low-energy excitations of the electronic subsystem, involves scalar and
gauge fields originating from the elastic properties and describing disorder
phenomena, like the distortions of the graphene lattice and structural
defects (see, for example, Ref. \cite{Jack07} and references therein). In
this setup, the spherical boundary models the edge of a negatively curved
graphene sheet. The boundary condition we have used ensures the zero flux of
a scalar field through the boundary. The Casimir effect in graphene-made
structures with zero curvature, like cylindrical and toroidal carbon
nanotubes, has been recently discussed in Ref. \cite{Bell09}.

\section{Acknowledgements}

A. A. S. was supported by the State Committee of Science Ministry of
Education and Science RA, within the frame of Grant No. SCS 13-1C040.

\appendix

\section{On the eigenvalues inside a sphere}

\label{sec:Zeros}

As it has been shown in section \ref{sec:WF}, the eigenvalues of the quantum
number $z$ in the region inside a spherical shell with the BC (\ref{Robin})
are zeros of the function $\bar{P}_{iz-1/2}^{-\mu }(u)$ for a given value of
$u>1$. The corresponding positive roots we have denoted by $z_{k}$, $%
k=1,2,\ldots $. From the relation $\bar{P}_{iz-1/2}^{-\mu }(u)=\bar{P}%
_{-iz-1/2}^{-\mu }(u)$ it follows that $z=-z_{k}$ are zeros of the function $%
\bar{P}_{iz-1/2}^{-\mu }(u)$ as well. Note that the latter can also be
written in the form%
\begin{equation}
\bar{P}_{iz-1/2}^{-\mu }(u)=(u^{2}-1)^{D/4}\tilde{p}_{iz-1/2}^{-\mu }(u),
\label{Pbartild}
\end{equation}%
where, for a given function $F(u)$ we have defined the notation
\begin{equation}
\tilde{F}(u)=AF(u)-\delta _{(j)}\frac{B}{a}\sqrt{u^{2}-1}\partial _{u}F(u),
\label{Ftilde}
\end{equation}%
and for the interior region $\delta _{(j)}=1$. From (\ref{Pbartild}) we see
that for $u>1$ the points $z=z_{k}$ are the zeros of the function $\tilde{p}%
_{iz-1/2}^{-\mu }(u)$. First of all let us show that these zeros are simple.

By taking into account that the function $R_{l}(r)=p_{iz-1/2}^{-\mu }(\cosh
r)$ is a solution of Eq. (\ref{Req}), the following integration formula can
be obtained%
\begin{eqnarray}
&&\int_{1}^{u}dx\,(x^{2}-1)^{D/2-1}[p_{iz-1/2}^{-\mu }(x)]^{2}=\frac{%
(u^{2}-1)^{D/2}}{2z}  \notag \\
&&\qquad \times \{[\partial _{z}p_{iz-1/2}^{-\mu }(u)]\partial
_{u}p_{iz-1/2}^{-\mu }(u)-p_{iz-1/2}^{-\mu }(u)\partial _{z}\partial
_{u}p_{iz-1/2}^{-\mu }(u)\}.  \label{IntFormp}
\end{eqnarray}%
From here, for $z=z_{k}$ one gets%
\begin{equation}
\int_{1}^{u}dx\,(x^{2}-1)^{D/2-1}[p_{iz-1/2}^{-\mu }(x)]^{2}=\frac{%
(u^{2}-1)^{(D-1)/2}}{2zB_{1}/a}p_{iz-1/2}^{-\mu }(u)\partial _{z}\tilde{p}%
_{iz-1/2}^{-\mu }(u)|_{z=z_{k}}.  \label{IntFormpk}
\end{equation}%
The left-hand side of this relation is positive (recall that $%
p_{iz-1/2}^{-\mu }(x)$ is a real function) and, hence, $\partial _{z}\tilde{p%
}_{iz-1/2}^{-\mu }(u)|_{z=z_{k}}\neq 0$. This shows that the zeros $z_{k}$
are simple.

The asymptotic expression for large positive zeros is found by using the
asymptotic formula%
\begin{equation}
P_{iz-1/2}^{-\mu }(u)\sim \sqrt{\frac{2}{\pi }}\frac{z^{-\mu -1/2}}{\sqrt{%
u^{2}-1}}\sin (z\,\mathrm{arccosh\,}u-\pi \mu /2+\pi /4),  \label{Pas}
\end{equation}%
for $z\gg 1$. From here we see that%
\begin{equation}
z_{k}\approx \pi \frac{2k+\mu +1/2-\delta _{0B}}{2\,\mathrm{arccosh\,}u}.
\label{zklarge}
\end{equation}

Now let us discuss possible purely imaginary zeros. With fixed $u$ and $l$,
the function $\bar{P}_{iz-1/2}^{-\mu }(u)$ has no purely imaginary zeros for
sufficiently small values of the ratio $\beta =Aa/B$. With increasing $\beta
$, started from some critical value $\beta _{l}^{(1)}(u)$, a pair of purely
imaginary zeros $z=\pm i\eta $, $\eta >0$, appears. The critical value $%
\beta _{l}^{(1)}(u)$ increases with increasing $l$ and, hence, the purely
imaginary zero first appears for the mode $l=0$ (see the left panel of
figure \ref{fig1}, where in the case $D=3$ the critical value is plotted
versus $r_{0}=\mathrm{arccosh\,}u$ for different values of $l$). By taking
into account that for $r\gg 1$ one has
\begin{equation}
P_{z-1/2}^{-\mu }(\cosh r)\approx \frac{\Gamma (z)e^{(z-1/2)r}}{\sqrt{\pi }%
\Gamma (1/2+z+\mu )},  \label{Plarger}
\end{equation}%
it is seen that for large values of $u$ we have $\eta \approx \beta +(D-1)/2$
and $\beta _{l}^{(1)}(u)\rightarrow -(D-1)/2$ for $u\rightarrow \infty $.
There are no purely imaginary zeros for $\beta <-(D-1)/2$. With the further
increase of $\beta >\beta _{l}^{(1)}(u)$ the value of $\eta $ increases and
for the second critical value $\beta =\beta _{l}^{(2)}(u)$ the energy of the
corresponding mode becomes zero. This corresponds to $\eta =z_{m}$. For $%
\beta >\beta _{l}^{(2)}(u)$ the energy of the mode becomes imaginary which
signals the instability of the vacuum.

\section{Summation formula}

\label{sec:SF}

In this appendix, by making use of the generalized Abel-Plana formula \cite%
{Sah1} (see also, \cite{Saha07Rev}), we derive a summation formula for the
series over the zeros of the function%
\begin{equation}
\bar{P}_{iz-1/2}^{-\mu }(u)=A(u)P_{iz-1/2}^{-\mu }(u)+B(u)\partial
_{u}P_{iz-1/2}^{-\mu }(u),  \label{PbarSum}
\end{equation}%
with respect to $z$, for given $u>1$ and $\mu \geqslant 0$. In general, the
functions $A(u)$ and $B(u)$ can be different from those in Eq. (\ref{AB}).
We shall denote the positive zeros, arranged in ascending order, by $z_{k}$,
$k=1,2,\ldots $, assuming that they are simple. The summation formula can be
obtained in a way similar to that used in Ref. \cite{Saha08} for the special
case $A(u)=1$ and $B(u)=0$ and we shall outline the main steps only (for the
summation formula over the zeros of the combination of the associated
Legendre functions of the first and second kinds see Ref. \cite{Saha09}).

We substitute in the generalized Abel-Plana formula%
\begin{eqnarray}
f(z) &=&\sinh (\pi z)h(z),  \notag \\
g(z) &=&\frac{e^{i\mu \pi }h(z)}{\pi i\bar{P}_{iz-1/2}^{-\mu }(u)}%
\sum_{j=\pm }\cos [\pi (\mu -jiz)]\bar{Q}_{jiz-1/2}^{-\mu }(u),  \label{fg}
\end{eqnarray}%
where the function $h(z)$ is analytic in the right-half plane of the complex
variable $z=x+iy$. The function $g(z)$ has simple poles at $z=z_{k}$. For
functions obeying the condition%
\begin{equation}
|h(z)|<\varepsilon (x)\exp (cy\,\mathrm{arccosh\,}u),\;|z|\rightarrow \infty
,  \label{Constr}
\end{equation}%
with $c<2$, $\varepsilon (x)e^{\pi x}\rightarrow 0$ for $x\rightarrow
+\infty $, the following formula is obtained
\begin{eqnarray}
&&\sum_{k=1}^{\infty }T_{\mu }(z_{k},u)h(z_{k})=\frac{e^{-i\mu \pi }}{2}%
\int_{0}^{\infty }dx\,\sinh (\pi x)h(x)  \notag \\
&&\qquad -\frac{1}{2\pi }\int_{0}^{\infty }dx\,\frac{\bar{Q}_{x-1/2}^{-\mu
}(u)}{\bar{P}_{x-1/2}^{-\mu }(u)}\cos [\pi (\mu +x)]\sum_{j=\pm }h(xe^{j\pi
i/2}).  \label{SumForm}
\end{eqnarray}%
Here we have introduced the notation%
\begin{equation}
T_{\mu }(z,u)=\frac{\bar{Q}_{iz-1/2}^{-\mu }(u)}{\partial _{z}\bar{P}%
_{iz-1/2}^{-\mu }(u)}\cos [\pi (\mu -iz)].  \label{T}
\end{equation}
Formula (\ref{SumForm}) is also valid for some functions having
branch-points on the imaginary axis, for example, in the case of functions
of the form $h(z)=F(z)/(z^{2}+z_{0}^{2})^{1/2}$, with $F(z)$ being an
analytic function. Note that, in the physical problem we have considered the
branch points correspond to $z=\pm iz_{m}$. Adding the corresponding residue
terms in the right-hand side, the formula (\ref{SumForm}) can be generalized
for functions $h(z)$ having poles in the right-half plane.

An equivalent expression for $T_{\mu }(z,u)$ is obtained by using the
relation
\begin{equation}
\bar{Q}_{iz-1/2}^{-\mu }(u)=\frac{B(u)e^{-i\mu \pi }\Gamma (iz-\mu +1/2)}{%
\Gamma (iz+\mu +1/2)(1-u^{2})P_{iz-1/2}^{-\mu }(u)},  \label{QW}
\end{equation}%
with $z=z_{k}$. This formula follows from the Wronskian relation for the
associated Legendre functions. Now, from Eq. (\ref{T}) for $z=z_{k}$ one gets%
\begin{equation}
T_{\mu }(z,u)=\frac{\pi e^{-i\mu \pi }B(u)|\Gamma (\mu +iz+1/2)|^{-2}}{%
(1-u^{2})P_{iz-1/2}^{-\mu }(u)\partial _{z}\bar{P}_{iz-1/2}^{-\mu }(u)}.
\label{T2}
\end{equation}

Formula (\ref{SumForm}) can be generalized for the case when the function $%
\bar{P}_{iz-1/2}^{-\mu }(u)$ has purely imaginary zeros at the points $z=\pm
iy_{k}$, $y_{k}>0$, $k=1,2,\ldots $, under the assumption that the function $%
h(z)$ obeys the condition%
\begin{equation}
h(z)=-h(ze^{-\pi i})+o(z-iy_{k}),\;z\rightarrow iy_{k}.  \label{Impolecond}
\end{equation}%
Assuming that the zeros are simple, in this case, on the right-hand side of
Eq. (\ref{SumForm}) we have to add the term
\begin{equation}
-i\sum_{k}\frac{\bar{Q}_{y-1/2}^{-\mu }(u)}{\partial _{y}\bar{P}%
_{y-1/2}^{-\mu }(u)}\cos [\pi (\mu -y)]h(iy)|_{y=y_{k}},  \label{Imzer}
\end{equation}%
at these poles and take the principal value of the second integral on the
right-hand side. The latter exists due to the condition (\ref{Impolecond}).
By using the relation (\ref{QW}) the term (\ref{Imzer}) can also be written
in the form
\begin{equation}
i\sum_{k}\frac{\pi B(u)e^{-i\mu \pi }}{P_{y-1/2}^{-\mu }(u)\partial _{y}\bar{%
P}_{y-1/2}^{-\mu }(u)}\frac{(u^{2}-1)^{-1}h(iy)}{\Gamma (\mu +y+1/2)\Gamma
(\mu -y+1/2)}|_{y=y_{k}}.  \label{Imzer2}
\end{equation}%
A physical example with purely imaginary modes using this result is
discussed in section \ref{sec:WF}.

\end{document}